\documentclass[notitlepage,aps,prd,reprint,groupedaddress,showpacs]{revtex4-1}

\usepackage{graphicx}
\usepackage{amsmath,amssymb, graphics, setspace}
\usepackage{commath}
\newcommand{\mathsym}[1]{{}}
\newcommand{\unicode}[1]{{}}

\begin{document}

\title{Data series subtraction  with  unknown and unmodeled background noise  }

\def\unitn{Department of Physics, University of Trento, and Trento Institute for Fundamental Physics and Applications, INFN,  38123 Povo, Trento, Italy}

\def\esac{European Space Astronomy Centre, European Space Agency, Camino bajo del Castillo s/n,
Urbanizaci\'on Villafranca del Castillo, Villanueva de la Can\~ada, 28692 Madrid, Spain}

\def\aei{Albert-Einstein-Institut, Max-Planck-Institut f\"ur Gravitationsphysik und Universit\"at Hannover,
Callinstrasse 38, 30167 Hannover, Germany}

\def\apc{APC, Universit\'e Paris Diderot, CNRS/IN2P3, CEA/Ifru, Observatoire de Paris, Sorbonne Paris Cit\'e, 10 Rue A.\,Domon et L.\,Duquet, 75205 Paris Cedex 13, France}

\def\ieec{Institut de Ci\`encies de l'Espai, (CSIC-IEEC), Facultat de Ci\`encies,
Campus UAB, Torre C-5, 08193 Bellaterra, Spain}

\def\ethz{ETH Z\"urich, Institut f\"ur Geophysik, Sonneggstrasse 5, 8092 Z\"urich}

\def\estec{European Space Technology Centre, European Space Agency, Keplerlaan 1, 2200 AG Noordwijk, The Netherlands}
\def\goddard{Gravitational Astrophysics Laboratory, NASA Goddard Space Flight Center}
\def\ox{Department of Physics, University of Oxford, Oxford OX1 3RH, UK}
\def\icl{High Energy Physics Group, Imperial College London. Blackett Laboratory, Prince Consort Road, London, SW7 2AZ}
% list of authors

\author{Stefano~Vitale}\email{stefano.vitale@unitn.it}\affiliation{\unitn}
\author{Giuseppe~Congedo}\altaffiliation[Current adrress: ]{\ox}\affiliation{\unitn}
\author{Rita~Dolesi}\affiliation{\unitn}
\author{Valerio~Ferroni}\affiliation{\unitn}
\author{Mauro~Hueller}\affiliation{\unitn}
\author{Daniele~Vetrugno}\affiliation{\unitn}
\author{William~Joseph~Weber}\affiliation{\unitn}
\author{Heather~Audley}\affiliation{\aei}
\author{Karsten~Danzmann}\affiliation{\aei}
\author{Ingo~Diepholz}\affiliation{\aei}
\author{Luigi~Ferraioli}\affiliation{\ethz}
\author{Ferran~Gibert}\affiliation{\ieec}
\author{Martin~Hewitson}\affiliation{\aei}
\author{Nikolaos~Karnesis}\affiliation{\ieec}
\author{Natalia~Korsakova}\affiliation{\aei}
\author{Miquel~Nofrarias}\affiliation{\ieec}
\author{Henri~Inchauspe}\affiliation{\apc}
\author{Eric~Plagnol}\affiliation{\apc}
\author{Oliver~Jennrich}\affiliation{\estec}
\author{Paul~W.~McNamara}\affiliation{\estec}
\author{Michele~Armano}\affiliation{\esac}
\author{James~Ira~Thorpe}\affiliation{\goddard}
\author{Peter~Wass}\affiliation{\icl}

\date{\today}

\begin{abstract}
LISA Pathfinder (LPF), the precursor mission to a gravitational wave  observatory of the European Space Agency, will measure the degree to which two test-masses can be put into free-fall, aiming to demonstrate a suppression of disturbance forces corresponding to a residual relative acceleration with a power spectral density (PSD) below ${\left(30\,\rm{fm/s^2/\sqrt{Hz}}\right)}^2$ around $1\,\rm{mHz}$.  In LPF data analysis, the disturbance forces are obtained as the difference between the acceleration data and a linear combination of other measured data series. In many circumstances, the coefficients for this linear combination are obtained by fitting these data series to the acceleration, and the disturbance forces appear then as the data series of the residuals of the fit.
Thus the background noise or, more precisely, its PSD, whose knowledge is needed to build up the likelihood function in ordinary maximum likelihood fitting, is here unknown, and its estimate constitutes instead one of the goals of the fit.
In this paper we present a fitting method that does not require the knowledge  of the PSD of the background noise.  The method is based on the analytical marginalisation of the posterior parameter probability density with respect to the  background noise PSD, and returns an estimate both for the fitting parameters and for the PSD. We show that both these estimates are unbiased, and that, when using averaged Welch's periodograms for the residuals,  the estimate of the PSD is consistent, as its error tends to zero with the inverse square root of the number of averaged periodograms.  Additionally, we find  that  the method is equivalent to some implementations of iteratively re-weighted least-squares fitting.  We have tested the method both on simulated data of known PSD, and on data from several experiments performed with the LISA Pathfinder end-to-end mission simulator.
\end{abstract}

\pacs{04.80.Nn,07.05.Kf,95.55.-n}

\maketitle

\section{\label{newintro} Introduction }

\begin{figure}[h]
\begin{center}
\includegraphics[width=0.48\textwidth]{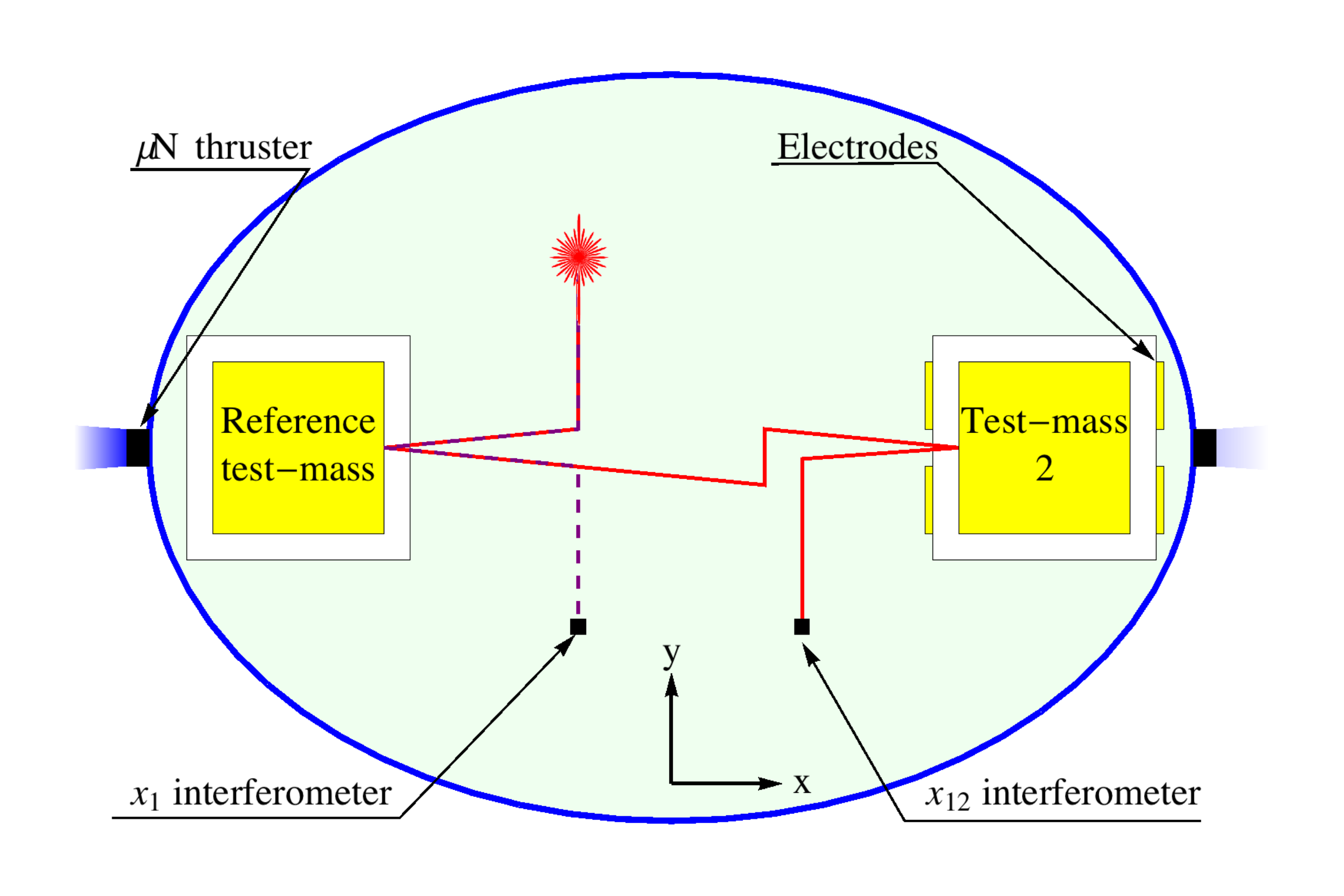}
\end{center}
\caption{Schematic of LPF. The figure shows the reference TM, TM2, and  the two laser interferometers --represented by their respective laser beam paths-- that measure $x_{1}$ and $x_{12}$ respectively. The $x$-axis, shown in the figure, is parallel to the line joining the centres of mass of the two TMs. The $z$-axis, normal to the figure, points toward the Sun. Also shown are the electrodes used to apply  the forces to TM2, necessary to keep it at nominally fixed distance from the reference TM. Similarly, the picture shows a pair of $\mu N$-thrusters that are used to force the satellite to stay at a nominally fixed position relative to the reference TM. Not shown in the figure are the electrodes and the  $\mu N$-thrusters used to control TM and satellite along degrees of freedom other than $x$. }

\label{lpf}   

\end{figure}

LISA Pathfinder (LPF)  \cite{LPF} is the  precursor mission to a gravitational wave (GW) observatory of the European Space Agency (ESA). Its primary goal is that of assessing if a set of reference test-masses (TMs) can be put into free motion, with residual accelerations, relative to the local inertial frame, having  a power spectral density (PSD)  less than 
${\left(30\,\rm{fm/s^2/\sqrt{Hz}}\right)}^2$, at frequencies between 1 and 30~mHz. 
This goal is pursued by measuring the relative acceleration of two  TMs, separated by  a nominal distance of 38 cm, along the line -- whose direction we call x -- joining their centres of mass (Fig. \ref{lpf}). The  relative motion between the TMs,  $x_{12}$, is measured by means of a laser interferometer, the output of which $s_{12}=x_{12}+n_{12}$ is affected by a  readout noise $n_{12}$  with  less than $\left(6\,\rm{pm/\sqrt{\mathrm{Hz}}}\right)^2$ PSD at mHz frequencies. 

The \emph{relative} acceleration  $a$ is then calculated by numerically performing the second time derivative \cite{deriv} of the interferometer output $s_{12}$:
\begin{equation}
\label{2deriv}
a=\frac{d^2 s_{12}}{dt^2}=\frac{d^2 x_{12}}{dt^2}+\frac{d^2 n_{12}}{dt^2}\equiv \frac{d^2 x_{12}}{dt^2}+a_r
\end{equation} 
where we have implicitly defined the readout acceleration noise $a_r$.

The TMs are not both free-falling along x. One TM, the inertial reference, is indeed following a pure geodesic orbit, but both the satellite, and the other TM, that we call TM2, are forced, by some force control loop, to stay nominally at fixed positions relative to the reference TM.  The satellite is actuated by a set of $\mu N$ thrusters within a feedback loop driven by the signal from a dedicated interferometer, which measures the relative displacement $x_{1}$ between the satellite and the reference TM. The second TM is instead subject to a weak electrostatic force commanded by  a feedback loop driven by the main interferometer signal $s_{12}$.

The relative motion of the satellite and the TMs, along degrees of freedom other than x, is also measured, either by laser interferometers or by capacitive sensing, and controlled by a combination of electrostatic forces and torques on the TMs, and of $\mu N$-thrusters-generated forces and torques acting on the satellite.

In standard operations, control loops keep the relative motion small enough that the system is expected to behave linearly, obeying  a set of linear dynamical equations \cite{Congedo}. For instance, the equation for $a$ is:

\begin{equation}
\label{dynamics}
\begin{split}
&a=\sum_{j}  R_{j}\divideontimes \frac{d^{2}s_{j}}{dt^{2}}  -\sum_{j}  \omega_{j}^{2} \divideontimes s_{j}+\sum_{j} A_{j}\divideontimes g_{j}^{c}+\\
&+g+a_{r}.
\end{split}
\end{equation}
The symbol $\divideontimes$ indicates time convolution.
$R_{j}$ in Eq. \ref{dynamics} is a linear operator that represents the unwanted pickup, by the differential interferometer, of generalised coordinates other than $x_{12}$, like for instance $x_1$. These coordinates are measured by the signals $s_j$, just as $s_{12}$ measures the coordinate $x_{12}$. Ideally $R_{j}$ should be zero, but imperfections and misalignments make it non-zero.   

In principle $R_{j}$ acts on coordinates, not on signals. Substituting coordinates with  signals, produces an extra term, as signals are always affected by some  readout noise. We absorb this term into the overall readout noise $a_r$. 

The readout noise,  because of the second time derivation, raises, in power, with the frequency $f$ as $\sim f^4$, and is expected to dominate the data above some 30 mHz, thus setting LPF's measurement bandwidth. 

The  generalised differential forces  \emph{per unit mass}, appearing on the right-hand side of Eq. \ref{dynamics} are split into three contributions:

\begin{itemize}
\item{The linear operator $\omega_{j}^{2}$  converts the relative motion of  the TMs and the satellite, along any of the degrees of freedom, into a differential force along x. For really free-falling TMs, $\omega_{j}^{2}$ should be zero. However static force gradients within the satellite makes the diagonal coefficients non-zero, while various kind of imperfections and misalignments contribute to non-diagonal terms.}
\item{The forces commanded by the control loops $g_{j}^{c}$, that are converted into true forces by the linear ``calibration'' operator $A_{j}$. $A_{j}$ should be just 1 when j corresponds to the electrostatic force commanded along $x$ on TM2,  and zero for any other value of $j$, but deviates from that because of imperfect calibration, delays and signal cross-talk.}
\item{The random forces $g$ stemming from all remaining disturbances, and whose measurement is the primary target of LPF.}
\end{itemize}

Measuring  $R_{j}$ and $A_{j}$ is one of the tasks of our analysis. Furthermore,  as the coupling of the TMs to the satellite  is expected to  be present also in a GW observatory like eLISA \cite{Vitale}, one of the goals of LPF is to give a measurement of $\omega_{j}^{2}$ to be compared with the prediction of the physical model of the system. 

\emph{The most important goal for LPF though, is that of measuring the PSD of  $g$, the parasitic forces that act on TMs and push them away from their geodesic trajectories.}

Eq. \ref{dynamics} suggests a natural way of achieving both these goals. Indeed both the $s_{j}$'s and the $g_{j}^{c}$'s are  known, as the first have been measured, and the second have been commanded by the control loops. Thus a fit of the $s_{j}$'s and the $g_{j}^{c}$'s to $a$, returns $R_{j}$,  $\omega_{j}^{2}$, and $A_{j}$, as best fit parameters, but  also allows the estimation of the PSD of $g$  from the fit residuals, that is from the difference between the acceleration data series and the fitting model. 

In reality we need to perform  such fits on the data from two  different kinds of experiment. 

When the target is that of measuring, with comparatively high  precision, the values  of  $R_{j}$, $\omega_{j}^{2}$ and $A_{j}$ (see Eq. \ref{dynamics}), 
we perform dedicated  calibration campaigns, where some proper guidance signals  are injected into the appropriate  control loops, so that the $s_{j}$'s and the $g_{j}^{c}$'s undergo large variations. This way  $R_{j}$, $\omega_{j}^{2}$ and $A_{j}$ can be measured with large Signal-to-Noise Ratio (SNR). 

When the target is instead a higher accuracy measurement of the PSD of the ultimate background acceleration noise, we do not apply any guidance signal, but just record acceleration noise data. These data are then fit to the $s_{j}$'s and the $g_{j}^{c}$'s, with the aim of separating  $g $ from the effect of the other force terms in the right hand side of Eq. \ref{dynamics}.  Indeed  $g $ becomes now  the residual of the fit, that is, the difference between the acceleration data and the best fit model. Actually, also other time series, like thermometer or magnetometer data, may  be fitted to the acceleration data to detect and separate specific disturbance sources.

It is worth stressing that, the $s_{j}$'s and the $g_{j}^{c}$'s cannot be turned off at any time so that an independent measurement of $g$ cannot be performed. A similar situation would also hold for a GW detector like eLISA, where large signals are expected to dominate the data at all times so that an independent measurement  of the background noise cannot be performed.

To perform  these fits we could not use a standard least squares method, and we had to develop a different fitting method. Indeed, to perform a least squares fit on data with coloured  background noise, as is certainly the case for LPF, one needs an {\it a priori} knowledge of the background noise PSD, either to set up a whitening filter, if the fit is performed in the time domain, or, for the more common case of  a fit in the frequency domain, to assign the statistical weights to each fit residual.  However, in our case,  the PSD is not known {\it a priori} and is actually one of the targeted outputs of  the fit. 

We have then developed a fitting method that works without an {\it a priori} knowledge of the background noise PSD. The method returns, besides the value of the fitting parameters, also an estimate for the background noise PSD. To achieve  a comparatively high precision  PSD estimation, the method preserves the ability of averaging over independent data stretches, like with  the standard Welch's averaged periodogram technique \cite{Welch}.
We use this method  in the framework of Bayesian estimation. However, we show in the paper that it can also be extended to the standard ``frequentist'' fitting approach. 

Over the last few years, different authors, in the framework of GW detection and Bayesian parameter estimation, have addressed the problem of fitting without a  complete \emph{a priori} knowledge of the noise PSD \cite{littenberg, Rover2007, Rover}. The emphasis of these studies was mostly on minimising the bias that  such a lack of knowledge may induce in the estimated signal parameters. This is a different target than the one we are discussing here, where the estimation of the noise is the main goal of the measurement, but the essence of the problem is the same. 

Two main approaches have been followed in these studies:

\begin{enumerate}

\item{Within the first approach \cite{Cornish, littenberg}, the value of the noise PSD $S_k\equiv S\left(f_k \right)$,  at each discrete frequency $f_k\equiv k/N T $, with N the length of the data series and T the sampling time, is  assumed to be described by some relatively smooth function of  frequency,  also depending on a vector of  some adjustable parameters $\vec{\eta}$,  $S_k=S\left(f_k,\vec{\eta}\right)$. 

The likelihood of the fit residuals becomes then a function  both  of  signal parameters and of $\vec{\eta}$.  

Appropriate prior probability densities -- often  some broad Gaussian or uniform  densities -- are then chosen for both the signal parameters and  $\vec{\eta}$.  Finally  the posterior likelihood for all parameters is numerically derived by   the Markov Chain Monte Carlo (MCMC) technique. 

Once the global likelihood has been derived, the marginal likelihood of the signal parameters alone, can be derived by numerically  marginalizing over the $\vec{\eta}$.}

\item{In the second approach \cite{Rover2007, Rover} the values of the PSD \emph{at each discrete frequency} $S_k$, are considered as independent parameters of the likelihood, each one distributed with a prior in the form of  a scaled inverse $\chi^2$ density, a family of distributions describing the statistics of the reciprocal of the square of Gaussian variables, that depend on two characteristic parameters \cite{Rover, GelmanBook}. This way the posterior density of both the signal parameters and of the $S_k$, becomes an analytical function of the observed residuals and of the prior parameters. Once the values for these prior parameters have been chosen for each frequency, and the authors discuss possible criteria for this choice, the likelihood can be calculated numerically by MCMC.}

\end{enumerate}

Our approach is close to the one in point 2 above with the following main differences and/or extensions:

\begin{enumerate}

\item{We adopt,  for the $S_k$, a family of priors that are uniform, either in the logarithm or in some small power  of $S_k$, over a wide, but finite range of values. These priors give a realistic representation of our knowledge on the residual noise of the system (see sect. \ref{Fitting}). The infinite range counterpart of these uniform priors can be obtained from the  scaled inverse $\chi^2$ family for  particular values of the prior parameters.} 
\item{With this assumption we are able to extend the method to the very important case where the time domain data are partitioned into (overlapping) stretches, so that the  standard Welch's averaged, and windowed,  periodogram of the residuals can be used for the fit \cite{Welch}. 

We show that, by using this approach,}
\begin{enumerate}
\item{The posterior likelihood can be \emph{analytically} marginalized over the $S_k$'s so that the marginalized likelihood of signal parameters, which takes a very simple form, can be easily  calculated numerically by MCMC, or numerically maximised within a standard fitting approach.}
\item{ $S_k$ can then be estimated analytically, and this estimate is shown to be consistent, itÕs error tending to zero as the inverse square root of the number of averaged periodograms.}
\item{The  above estimate shows a slight  bias that depends on the specific prior adopted, but  this bias tends to zero  linearly with the inverse of the number of averaged periodograms.}
\end{enumerate}

\end{enumerate}

The paper presents such a method and is organised as follows. In Sec. \ref{sec:2} we describe the method. In Sec. \ref{test} we give a test of the method with synthetic data of known PSD, and  we present a few examples of its application to the reduction of data from LPF end-to-end mission simulator. Finally, in Sec. \ref{disc} we briefly discuss the results and the possibility of extending the method to signal extraction for the data of  GW detectors.

\section{\label{sec:2} Maximum likelihood  with unknown coloured noise }

Though the method is general, for the sake of clarity  we will continue to refer to the example of LPF. The main signal for LPF is the relative acceleration data series $a\sbr{n}$. We assume that the acceleration data series may be modelled as

\begin{equation}
\label{eq:2.1}
a\left[n\right]=g_m\left[ n,\vec{\theta}  \right]+ g\left[ n \right]
\end{equation}

In Eq. \ref{eq:2.1} the data series  $g_m \left[ n, \vec{\theta}\right]$ consists of  the samples of a linear combinations of measured signals, like for example $\omega_{j}^{2}s_{j}$ or $A_{j}g_{j}^{c}$ in Eq. \ref{dynamics}. $g_m \left[ n, \vec{\theta}\right] $ may have been obtained by processing those signals   by some algorithm that depends on a set of parameters $\theta_{i}$  that we have arranged into the vector $\vec{\theta}$. For instance, the  force commanded on TM2 along x is described by $A  g_2^c \left( t - \tau_2 \right)$, where $ g_2^c \left( t \right) $  is the recorded commanded force. Thus $\vec{\theta}$ contains in this case the calibration amplitude $A$ (nominally one) and the delay $\tau_2$, with the effect of the latter calculated by numerical interpolation of the data. 

$g\left[ n \right]$ represents the residual differential force noise time series, the main objective of the measurement, though corrupted by the superposition  of the time series of the readout noise $a_r\sbr{n}$. As discussed before, Eq. \ref{eq:2.1} indicates that   $g\sbr{n}$  might be derived from  $g\sbr{n,\vec{\theta}}\equiv a\left[n\right]-g_m\left[ n,\vec{\theta}  \right]$, the  time series of ``residuals'', a series that depends parametrically on  $\vec{\theta}$.

The equality in Eq. \ref{eq:2.1} is preserved by moving to the Fourier domain so that one can write: 

\begin{equation}
\label{eq:2.2}
 \tilde{g}\left[ k,\vec{\theta} \right] =\tilde{a}\left[k\right]-\tilde{g}_m\left[ k,\vec{\theta}  \right] 
\end{equation}
where the tilde indicates a Discrete Fourier Transform (DFT). 

We define the  DFT of a stretch of $N$ data of any  series $y\left[n\right]$ as

\begin{equation}
\label{eq:DFT}
 {\tilde{y}}\left[ k \right]=\frac{1}{\sqrt{N}}\sum\limits_{n=0}^{N-1}{ {y}\left[ n \right]w\left[ n \right]{{e}^{-i\ n\ k\frac{2\pi }{N}}}}
\end{equation}

In the transformation we have already included  the multiplication of data by a  properly selected spectral window $w\left[ n \right]$. This is common practice in spectral estimation  to avoid excess spectral leakage \cite{percival}.

\subsection{\label{sec:2.1} Building up the likelihood function }

We now want to discuss the joint probability density  function of the  residuals $\tilde{g}\left[ k,\vec{\theta}\right]$, conditional to a specific choice of the values of the  $\theta's$. We will assume here that all noise sources are Gaussian and zero mean, so that also the $\tilde{g}\left[ k,\vec{\theta}\right]$ are  zero-mean and Gaussian. Large non Gaussian noise in LPF, like glitches and spikes, can satisfactorily be treated as being constituted of signals and subtracted from the data with minimal corruption of the data, given the focus on the lowest frequencies.

The  residuals $\tilde{g}\left[ k \right]$ are complex quantities.  It is shown in  appendix Sec. \ref{Appendix} that, for $\abs{k}\ge k_{o}$ and $\abs{k-k'}\ge k_{1}$, where $k_{o}$ and $k_{1}$ are  integers depending on the adopted spectral window, $Re\lbrace\tilde{g}\left[ k \right]\rbrace$,  $Im\lbrace\tilde{g}\left[ k \right]\rbrace$, $Re\lbrace\tilde{g}\left[ k' \right]\rbrace$,  and $Im\lbrace\tilde{g}\left[ k \right]\rbrace$ may be considered, with good accuracy, as all zero-mean, independent Gaussian variables. For instance, for the Blackman-Harris spectral window we commonly use in LPF data analysis, we show in appendix \ref{Appendix}, that a safe assumption is $k_{o}=k_{1}=4$. The variances of  $Re\lbrace\tilde{g}\left[ k \right]\rbrace$,  and of $Im\lbrace\tilde{g}\left[ k \right]\rbrace$, are given by (see appendix Sec. \ref{Appendix}):
\begin{equation}
\label{eq:variance}
\sigma^{2}_{Re\left\lbrace\tilde{g}\left[ k \right]\right\rbrace}=\sigma^{2}_{Im\lbrace\tilde{g}\left[ k \right]\rbrace}=\frac{1}{2}S_{k},
\end{equation}
where  $ S_{k}$ is the frequency  averaged discrete time PSD of $g\left[n\right] $,  at the frequency $f_k=k/N T$ defined as:
\begin{equation}
\label{eq:Sk}
\begin{split}
S_{k}=\frac{1}{2\pi}\int_{-\pi }^{\pi }\tilde{S}_{\tilde{g}}\left(\phi\right){\abs{w\left(\phi-k\frac{2\pi}{N}\right)}}^{2}d\phi.
\end{split}
\end{equation}

Here $w\left(\phi \right)$ is the discrete-time Fourier transform of $w\left[n\right]$ and $\tilde{S}_{\tilde{g}} \left(\phi\right)$ is the discrete-time power spectral density of the \emph{infinite length} $g\left[ n \right] $ series,  from which the set of $N$ data under analysis has been extracted. Notice that, if aliasing is avoided,  then $\tilde{S}_{\tilde{g} }\left(\phi\right)=T S\left(f=\phi/ 2 \pi T\right)$,  where $S(f)$ is the ordinary PSD of the continuous process $g\left(t\right)$ of which  $g\left[ n \right]=g\left(t=nT\right) $ constitute the series of the samples. 

Under  the hypotheses above, the joint conditional probability density function of  $Re\lbrace\tilde{g}\left[ k \right]\rbrace$ and  $Im\lbrace\tilde{g}\left[ k \right]\rbrace$ is given by: \begin{equation}
\label{eq:Gaussian probability}
\begin{split}
P\left(\vec{g}\middle \vert  \vec{\theta} , \vec{S}\right)=\prod_{k\in Q}\frac{1}{\pi S_{k}}e^{-\frac{{\left\rvert \tilde{g}\left[ k, \vec{\theta}\right]\right\lvert}^{2}}{S_{k}}}.
\end{split}
\end{equation}
In Eq. \ref{eq:Gaussian probability}, Q is the subset $Q=\set{k_{0}, k_{0}+k_{1}, k_{0}+2 k_{1}, ....}$, of the integer set $0 \le k \le N/2 $. In the same equation, and  in the rest of the paper, we have organised the residuals $\tilde{g}\left[ k,\vec{\theta}\right]$ and the $S_{k}$, with $k\in Q$, into the vectors $\vec{g}$ and  $\vec{S}$ respectively.

It is standard, in spectral analysis, to partition data series into shorter stretches, and to average the spectral estimate over these stretches. Different stretches are treated as statistically independent, even in the presence of partial overlap between adjoining stretches, if these have been tapered at their ends  with a proper spectral window.
Assuming that the data series have been partitioned into $N_{s}$ independent stretches, the probability density in Eq. \ref{eq:Gaussian probability} becomes:
\begin{equation}
\label{eq:averaged probability}
\begin{split}
P\left(\overline{\vec{g}}\middle\vert  \vec{\theta},\vec{S} \right)=\prod_{k\in Q}\frac{1}{{\left(\pi S_{k}\right)}^{N_{s}}}e^{-N_{s}\frac{\overline{{\left\rvert \tilde{g}\left[ k,\vec{\theta} \right]\right\lvert}^{2}}}{S_{k}}},
\end{split}
\end{equation}

where the bar represents an average over the $N_{s}$ stretches.

In Eq. \ref{eq:averaged probability}  we have written the probability density of the data as also being conditional on  $\vec{S}$. In a standard fitting procedure, these coefficients are assumed to be known. On the contrary, here we discuss the case where the components of  $\vec{S}$  are unknown and must be estimated from the data. 

We now write down the posterior probability density of $\vec{S}$ and $\vec{\theta}$.

\begin{equation}
\label{eq:2.4}
\begin{split}
&P \left(\vec{\theta}, \vec{S} \middle \vert \overline{\vec{g}}\right)=\\&\frac{P\left(\overline{\vec{g}} \middle \vert  \vec{\theta},\vec{S} \right)\times \prod_{k \in Q}P\left(S_{k}  \right)\times P\left(\vec{\theta}\right)}{\int P\left(\overline{\vec{g}} \middle \vert  \vec{\theta},\vec{S} \right)\times \prod_{k\in Q}P\left(S_{k}  \right)d S_{k}\times P\left(\vec{\theta}\right)d\vec{\theta}}.
\end{split}
\end{equation}

In Eq. \ref{eq:2.4} we have made the key assumptions that $S_{i}$ is independent of $S_{j}$ if $i \ne j$, and that both are independent of $\vec{\theta }$. This way the joint \emph{prior} probability density $P\left(\vec{\theta},\vec{S}\right)$ splits into the product  of  the separated prior probability densities  $P\left(\vec{\theta}\right)$ and  $P\left(S_{k}\right)$. 

While the independence of $\vec{\theta}$ and $\vec{S}$ is rather natural, the physical basis for the independence of $S_i$ and $S_j$ , when $i \ne j$, may need some justification. 

Our {\it a priori}  knowledge of the noise PSD , in the case of an instrument where the signal cannot be 'turned off' and the noise independently measured, is rather limited. Unexpected lines could be present in the spectrum, such that nearby coefficients, $S_i$ and $S_{i \pm1}$, may differ even in order of magnitude. 
Thus even the perfect knowledge of  $S_i$ would not give us any significant information on the probability density function  of any of the other $S_j$'s, which is the very definition of independent random variables.

\subsection{\label{Fitting} Fitting}

We now discuss the use of the likelihood function in Eq. \ref{eq:2.4} for the purpose of fitting. As most of our fits involve nonlinear functions of parameters, our preferred approach is that of Bayesian parameter estimation with the  Markov Chain Monte Carlo (MCMC) method, but we will also consider a more conventional approach wherein one searches for a maximum of the likelihood as a function of fitting parameters. We will show now that, whatever the selected approach, with a proper choice of the prior probability density of the components of $\vec{S}$, the parameter space can be reduced to just that  of  $\vec{\theta}$.

In order to do this, we assume that the prior of $S_k$ is  uniform as a function of either some small power of $S_{k}$, or of $\log{(S_{k})}$, between two values $S_{k,a} \ll S_{k,b}$. This is not exactly the same as using  Jeffrey's non-informative prior \cite{Jeffreys}, though it may be approximated by this under some assumptions that we will make later.  

Such a choice for the prior, in particular a uniform density as a function of $\log{(S_{k})}$, again closely reflects our {\it a priori} physical knowledge of the noise. Indeed, as said,  in LPF, as in a GW observatory like eLISA, signals cannot be turned off and the background noise cannot be independently measured.  Though we may have theoretical  models for the sources of $g$,  unexpected noise sources may lead to a PSD deviating  from theoretical expectations  even in order of magnitude. 

With this choice of the prior, $P\left(\vec{\theta},\vec{S} \middle \vert \overline{\vec{g}}\right)$ can be analytically integrated over the $\vec{S}$ space, to obtain a marginalized likelihood which is only a function of  $\vec{\theta}$. By assuming that $P\left(\vec{\theta}\right)$ is bound to a domain on which $\overline{{\left \lvert \tilde{g}\left[ k\right]\right \rvert }^{2}}\ll S_{k,b}$,  and that, for any value of $\vec{\theta}$,  $S_{k,a} \ll \overline{{\left \lvert \tilde{g}\left[ k\right]\right \rvert }^{2}}$, the integration over $S_{k}$ can be extended from zero to infinity:

\begin{equation}
\begin{split}
\label{defmarg}
&P_{marg} \left(\vec{\theta} \middle \vert \overline{\vec{g}} \right)=\int_{0}^{\infty} P\left(\vec{\theta},\vec{S} \middle \vert \overline{\vec{g}}\right)d \vec{S}.
\end{split}
\end{equation}

By performing the integration we obtain:

\begin{equation}
\begin{split}
\label{lmarg}
P_{marg} \left(\vec{\theta} \middle \vert \overline{\vec{g}} \right)=\frac{\displaystyle \prod_{k\in Q}\left(\overline{{\left\lvert \tilde{g}\left[ k,\vec{\theta}\right] \right \rvert }^{2}}\right)^{m-N_{s}}}{\displaystyle \int\prod_{j\in Q}\left(\overline{{\left\lvert \tilde{g}\left[ j,\vec{\theta'}\right] \right \rvert }^{2}}\right)^{m-N_{s}}P\left(\vec{\theta'}\right)d\vec{\theta'}},
\end{split}
\end{equation}
for a uniform prior either in $S_{k}^{m}$ or in $log{\left(S_{k}\right)}$, in which case  one should put $m=0$ in the formulas. 

In addition, the likelihood  $P\left(\vec{\theta},\vec{S}  \middle \vert \overline{\vec{g}}\right)$, for  any given value of $\vec{\theta}$, reaches a maximum $P_{max} \left(\vec{\theta} \middle \vert \overline{\vec{g}} \right)$ for some values   $S_{k,max} $ of the $S_{k}$'s. Both the value of  $P_{max} \left(\vec{\theta} \middle \vert \overline{\vec{g}} \right)$, and of $S_{k,max} $  can be calculated analytically by 
differentiation of Eq. \ref{eq:averaged probability}. We get

\begin{equation}
\begin{split}
\label{sbest}
S_{k,max}=\frac{N_{s}}{N_{s}-m+1}\overline{{\left\lvert \tilde{g}\left[ k,\vec{\theta}\right] \right \rvert }^{2}},
\end{split}
\end{equation}

and

\begin{equation}
\begin{split}
\label{lmax}
&P_{max} \left(\vec{\theta} \middle \vert \overline{\vec{g}} \right)=\\&=\frac{{{\text{e}}^{-\left( {{N}_{s}}+1 \right)}}{{\left( {{N}_{s}}+1 \right)}^{{{N}_{s}}+1-m}}}{{{N}_{s}}\left( {{N}_{s}}-m-1 \right)!} \frac{P_{marg} \left(\vec{\theta} \middle \vert \overline{\vec{g}} \right) }{\displaystyle \prod_{k\in Q}\overline{{\left\lvert \tilde{g}\left[ k,\vec{\theta}\right] \right \rvert }^{2}}}.
\end{split}
\end{equation}
We note that  Eq.~\ref{sbest}, in the limit of large $N_s$, where all priors give the same formula, anticipates our result for the estimate of $S_k$  that we further discuss in detail in subsection \ref{method}. 

These results allow us to restrict the fitting to the $\vec{\theta}$ parameter space. In the maximisation approach, one can  search for a maximum of $P_{max} \left(\vec{\theta} \middle \vert \overline{\vec{g}}\right)$, that will also be a maximum of $P \left(\vec{\theta},\vec{S} \middle \vert \overline{\vec{g}} \right)$.
Within the Bayesian approach, the likelihood mapping  may be performed   over just   $\vec{\theta}$, by using $P_{marg} \left(\vec{\theta} \middle \vert \overline{\vec{g}} \right)$. 

In both cases, the results in  Eqs. \ref{lmarg}, \ref{sbest}, and \ref{lmax}  indicate that, for large enough $N_{s}$ the logarithm of the likelihood to be either maximised, or used in MCMC mapping, is

\begin{equation}
\label{eq:master formula}
\begin{split}
&\Lambda \left( \vec{\theta}\right)\equiv\log \left( P_{marg}\left(\vec{\theta} \middle \vert \overline{\vec{g}}\right)  \right) =\\&=-N_{s} \sum_{k \in Q}  \log \left( \overline{{\rvert \tilde{g}\left[ k,\vec{\theta} \right]\lvert}^{2}}   \right)+C,
\end{split}
\end{equation}
instead of the standard least squares fitting result with known $S_k$, 
\begin{equation}
\label{eq:lsalt}
\begin{split}
\Lambda \left( \vec{\theta}\right) =-N_{s} \sum_{k \in Q}  \frac{ \overline{{\rvert \tilde{g}\left[ k,\vec{\theta} \right]\lvert}^{2}}} {S_{k}}+C'.
\end{split}
\end{equation}
Here, $C$ and $C'$ are just constants.   In essence, according to Eq. \ref{eq:master formula}, in the presence of unknown and unmodeled noise, any fit must minimise not the mean square residuals, but rather the sum of their logarithm.   This is one of the main results of the paper. 

As already mentioned in the introduction, a likelihood proportional to that in Eq. \ref{lmarg} has been found for the special case $N_s$~=~1 and $m$~=~0 by \cite{Rover}. Our result generalises it to the experimentally important case of averaged and windowed periodograms, and to the case of  $m \le 0$.

For the rest of the paper we will call  the likelihood in Eq. \ref{eq:master formula}, the ``logarithmic'' likelihood (LL).

\subsection{\label{sec:2.4} Truly independent DFT coefficients and spectral resolution}

In order to maintain the accuracy of the result in Eq.  \ref{eq:master formula}, one should fulfil the condition  $\abs{k-k'}\ge k_{1}$, dropping a great number of DFT coefficients, and thus lowering the spectral resolution of the fit. This might be undesirable at the lowest frequencies,  where the relative spectral resolution is rather low. The inaccuracy deriving from summing over \emph{all} DFT coefficients in Eq. \ref{eq:master formula} reduces to  counting  each independent DFT coefficient more than once. Thus, using  such a likelihood function may overestimate the number of degrees of freedom of the fit and may lead to an underestimate of the parameter errors. This effect may  be corrected  for by scaling down the likelihood by an appropriate correction factor $\gamma$, that is by using the likelihood $\tilde{\Lambda}\left( \vec{\theta}\right)\equiv \gamma\times\Lambda \left( \vec{\theta}\right)$.  This is further discussed in Sec. \ref{test}.

As for  the coefficients  with $k < k_{0}$,  in spectral estimation these are   discarded  in any case, because of the strong bias they suffer from the leakage of spectral power from frequencies below $f=1/N T$.

\subsection{\label{method} Summary of fitting procedure  and the uncertainty on parameter estimates}

The derivation in the preceding sections  leads to a very simple implementation of the method. The entire machinery of a standard frequency domain fit can be retained, provided that the usual likelihood function is replaced with that in Eq. \ref{eq:master formula}.

Thus, for instance, within the framework of Bayesan estimation, rather than sampling from the joint likelihood of $\vec{\theta}$ and $\vec{S}$, parameter estimates of $\vec{\theta}$ are obtained from the marginalised LL likelihood, which can be numerically mapped over the space of the $\theta$'s by  standard MCMC. 

Such a calculation does not return an estimate for the $S_{k}$ and their errors. These however can be estimated as follows. 
Let's write the conditional mean value of any integer power of $S_{j}$, $S_{j}^{m}$ as

\begin{equation}
\label{eq:momenta}
\begin{split}
\left \langle S_{j}^{m} \middle \vert \overline{\vec{g}}  \right \rangle=\frac{
\displaystyle\int_{}^{}P\left(\vec{\theta}\right)d\vec{\theta}\int_{0}^{\infty} S_{j}^{m}P\left(\overline{\vec{g}}\middle\vert\vec{\theta},\vec{S}\right)P\left(\vec{S}\right)d\vec{S}}{\displaystyle\int_{}^{}P\left(\vec{\theta}\right)d\vec{\theta}\int_{0}^{\infty} P\left(\overline{\vec{g}}\middle\vert\vec{\theta},\vec{S}\right)P\left(\vec{S}\right)d\vec{S}}
\end{split}
\end{equation}

This integral may be somewhat reduced if one uses  the prior  for $\vec{S}$ discussed above. Indeed one can check (see the appendix \ref{Appendix2}) that

\begin{equation}
\label{eq:sands1}
\begin{split}
\left \langle S_{j} \middle \vert \overline{\vec{g}}  \right \rangle=\frac{N_{s}}{N_{s}-m-1}\left \langle \overline{{\left\vert\tilde{g}\sbr{j}\right\vert}^{2}}\right \rangle\\
\end{split}
\end{equation}

and that 

\begin{equation}
\label{eq:sands2}
\begin{split}
\left \langle S_{j}^{2} \middle \vert \overline{\vec{g}}  \right \rangle=\frac{N_{s}^{2}}{\left(N_{s}-m-1\right)\left(N_{s}-m-2\right)}\left \langle \overline{{\left\vert\tilde{g}\sbr{j}\right\vert}^{4}}\right \rangle.\\
\end{split}
\end{equation}

Here the mean value $\left \langle \overline{{\left\vert\tilde{g}\sbr{j}\right\vert}^{4}}\right \rangle$ is the value of $ \overline{{\left\vert\tilde{g}\sbr{j,\vec{\theta}}\right\vert}^{4}}$ averaged over the total posterior probability density of $\vec{\theta}$. Thus, once this posterior probability has been calculated as described above, in principle also the values of $S_j$ and of $S_j^2$, and then of the variance of $S_j$, can also be numerically calculated.

The numerical calculation can be avoided if  the parameter posterior distribution is symmetric enough around the mean values, a condition  we always find  satisfied in our numerical calculations. In this case the mean values in the rightmost terms of the eq. \ref{eq:sands1} and eq. \ref{eq:sands2} will coincide  with the values they take  at the mean values $\vec{\theta}=\vec{\theta}_{o}$ where the likelihood is also a maximum. Then, for $N_{s}\gg 1,m$:

\begin{equation}
 \label{sigma}
\begin{split}
&\left \langle S_{j} \middle \vert \overline{\vec{g}}  \right \rangle\simeq \overline{{\left\vert\tilde{g}\sbr{j,\vec{\theta}_{o}} \right\vert}^{2}}\\
&\sigma_{S_{j}}\equiv \sqrt{\left\langle S_{j}^{2} \middle \vert \overline{\vec{g}}  \right\rangle-{\left\langle S_{j} \middle \vert \overline{\vec{g}}  \right\rangle}^{2}}\simeq\frac{\overline{{\left\vert\tilde{g}\sbr{j,\vec{\theta}_{o}} \right\vert}^{2}}}{\sqrt{N_{s}}}.
\end{split}
\end{equation}

Eq. \ref{sigma} shows that the estimate of the $S_j$ is consistent, as its error decreases as $1/\sqrt{N_s}$.

It is worth noticing that, by increasing  $N_s$, besides improving the precision of the noise estimate, one also rapidly suppresses any residual dependance of both the likelihood in   Eq. \ref{lmarg} and the momenta in Eqs. \ref{eq:sands1} and \ref{eq:sands2}, on $m$, the only remaining parameter  in our approach that is still needed to characterise the noise.  Indeed marginalisation  removes the dependance of the likelihood on  the $S_k$'s but does not remove its dependence on  parameters, like $m$,  entering in the prior for $S_k$, whose value must then be decided in advance (see for instance ref \cite{Rover}). It appears then that averaging  over many periodograms may be an effective way to  remove such a residual bias.

In this sense,  neither the estimate for $S_k$ nor that for $\vec{\theta}$ are  biased by  noise modelling.

Thus, in summary, the fitting proceeds in two steps: after having mapped the marginalised  likelihood LL to estimate $\vec{\theta}$,  $\vec{S}$ and their uncertainties can be evaluated, through eq. \ref{sigma}, from the fit residuals evaluated for  $\vec{\theta}=\vec{\theta}_{o}$, where LL takes its maximum value.

\subsection{\label{irls} Relation to the Iteratively Re-weighted Least Squares (IRLS) method}

A popular method  to perform least square fitting without an {\it a priori} knowledge of the expected residual PSD, is that of performing the fit iteratively. The procedure starts by performing an ordinary least squares fit, maximising then the  likelihood in eq. \ref{eq:lsalt},  by using some arbitrary initial value for the $S_{k}$, like for instance $S_{k}=1$, or alternatively by using the spectrum of the raw data before fitting. Once the maximum has been found at some point $\vec{\theta}_{o}$, the square modulus of the residuals of this first fit, $\overline{{\left\vert\tilde{g}\sbr{k,\vec{\theta}_{o}} \right\vert}^{2}}$  are used  as the  $S_{k}$'s in a new run of the fit. Often the residuals are smoothed over the frequency, or fitted to some smooth model before using them as weights in the next iteration of the fit. However the use of the residuals as they are is unbiased and numerically lighter, thus we prefer  to restrain the discussion to just this case. The procedure is then iterated until the residuals of the fit and  the $S_{k}$ do not change anymore to within some tolerance. The procedure usually converges quite rapidly. We will show here that the parameter values that are obtained with this iterative procedure are the same as those obtained by a LL.

Consider that, at the n\textsuperscript{th} step of the procedure, one wants to minimise
\begin{equation}
\begin{split}
\label{ratio}
\chi_{n}^{2}\equiv N_{s}\sum_{k \in Q}\frac{\overline{{\left\vert\tilde{g}\sbr{k,\vec{\theta}_{n}} \right\vert}^{2}}}{\overline{{\left\vert\tilde{g}\sbr{k,\vec{\theta}_{n-1}} \right\vert}^{2}}}
\end{split}
\end{equation}

as  function of $\vec{\theta}$. If the method converges, at some point the sequence becomes stationary and independent of n. Let's write 

\begin{equation}
\begin{split}
\label{expand}
&\overline{{\left\vert\tilde{g}\sbr{k,\vec{\theta}_{n}} \right\vert}^{2}}\simeq \overline{{\left\vert\tilde{g}\sbr{k,\vec{\theta}_{n-1}} \right\vert}^{2}}+\\
&\sum_{i} \frac{\partial \overline{{\left\vert\tilde{g}\sbr{k,\vec{\theta}_{n-1}} \right\vert}^{2}}}{\partial \theta_{i}}\delta\theta_{i}
\end{split}
\end{equation}

substituting in Eq. \ref{ratio}, we get
\begin{equation}
\begin{split}
\label{stationary}
\chi_{n}^{2}&\simeq N_{s}\sum_{k \in Q}\left(1+ \sum_{i} \frac{\partial \overline{{\left\vert\tilde{g}\sbr{k,\vec{\theta}_{n-1}} \right\vert}^{2}}/\partial \theta_{i}}{\overline{{\left\vert\tilde{g}\sbr{k,\vec{\theta}_{n-1}} \right\vert}^{2}}}\delta\theta_{i}  \right)\\&=N_{Q} N_{s}+\sum_{i}\frac{\partial N_{s}\sum_{k \in Q}\log\left( \overline{{\left\vert\tilde{g}\sbr{k,\vec{\theta}_{n-1}} \right\vert}^{2}}\right)}{\partial \theta_{i}}\delta\theta_{i}.
\end{split}
\end{equation}

Where $N_{Q}$ is the number of elements in Q. If the sequence $\chi_{n}^{2}$ must become stationary, then from some value of $n$ on, the derivatives in the right hand side of Eq. \ref{stationary} must become zero. Then  $N_{s}\sum_{k \in Q}\log\left( \overline{{\left\vert\tilde{g}\sbr{k,\vec{\theta}_{n-1}} \right\vert}^{2}}\right)$ reaches a maximum and hence so does the likelihood in Eq. \ref{eq:master formula}.

This shows that IRLS and LL achieve the same estimate for the best fit parameters. This will be also shown numerically in Sec. \ref{test}.

For non-linear fits, when the maximum of the likelihood  must be searched for numerically, the straightforward maximisation of the likelihood in  Eq. \ref{eq:master formula} is substantially quicker than the IRLS method, as the maximisation with respect to the parameters must be performed only once. On the other hand, when the dependence of the likelihood on the parameters is linear, that is if the $\theta$'s are just amplitude coefficients, then the maximisation, at each step,  reduces to solving a system of linear  equations  and the IRLS method may be  substantially faster.

A theoretical estimate of  parameter errors with the IRLS method, is not straightforward. For nonlinear fit models, once the stationary solution has been reached, one can use the residuals from this solution in place of the $S_{k}$ in the least squares likelihood of Eq. \ref{eq:lsalt}, and perform an MCMC map, from which the parameter errors can be calculated. If the fit model is linear, one can apply the standard error propagation, or Fisher matrix technique, to the stationary solution and derive the errors from that. Though this is an heuristic approach, we will show in Sec. \ref{test}, that, at least in the case of the non-linear models that we have studied in the present paper, it gives results consistent with those obtained with the LL fits.

Finally we want to make a couple of remarks:
\begin{itemize}
\item{The equivalence of the two methods shows that an IRLS fit that uses the single DFT coefficients from one fit to weight the next fit -- as opposed to taking some smooth, few parameter model of the spectrum frequency dependence, such as a polynomial -- implicitly assumes that the $S_{k}$ are uncorrelated and uniformly distributed in some logarithmic or power law space, as our LL method explicitly does.}
\item{The assumption of uncorrelated and uniformly distributed $S_{k}$  allows for arbitrary variations of the PSD from one frequency to the next one and allows us then to weight neighbouring frequencies quite differently based on the DFT of their residuals. The impact of this inhomogeneous weighting is limited in practice by averaging the spectrum over many time windows, which reduces the fluctuations in the mean square residuals between neighbouring bins, and by using a small number of disturbance parameters $\vec{\theta}$. While we see no evidence of anomalies in the results of our analyses, we bear in mind that any assumption about the $S_{k}$ can have an impact on the analysis. Our technique allows for an inhomogeneous weighting of points,  as is consistent with our hypothesis of uncorrelated and uniformly distributed $S_{k}$ , and, unlike a low-order polynomial model of the spectrum frequency dependence, allows for lines, ÒbumpsÓ, and other unexpected spectral features.}
\end{itemize}

\section{\label{test} Application to  data}

We have tested the method in two ways:

\begin{itemize}
\item{With the aim of  verifying its accuracy, we have performed a noise decomposition  exercise on  a set of data generated from a  fully known PSD.}
\item{In order to test the practical usability of the method with non ideal data, we have used it in various data reduction exercises on the data from the LPF end-to-end mission simulator.}
\end{itemize}

In the next sections we describe these tests and report on their results, while in Sec. \ref{disc} we discuss their significance.

\subsection{\label{simul} Test on data with a known PSD}

To perform this test we have generated  some data series with  spectral content and dynamic range similar to those of LPF data. 
A base data series  has been created which is the sum of two components. The first, $a\sbr{n}$, simulating the acceleration data, is the numerical second derivative, obtained as described in \cite{deriv}, of  a ``red'' noise series. This red series is   composed of  a $1/f^6$ PSD low frequency tail merging, at 10 mHz, into a flat PSD. 
The second series, $g^{c}\sbr{n}$, which simulates the feedback force data, consists of a  tail with  $1/f^4$ PSD merging into a flat plateau at 0.3 mHz.
This base series is contaminated by independently superposing a  series $\alpha_{1,o}\delta a\sbr{n}$ to $a\sbr{n}$ \emph{before} derivation and one $\alpha_{2,o}\delta g\sbr{n}$ to $g^{c}\sbr{n}$.  $\delta a\sbr{n}$ has  a $1/f^6$ PSD, while  $\delta g\sbr{n}$ has a band-pass structure with a broad peak around 2 mHz. Thus the data consist of

\begin{equation}
\label{totaldata}
g\sbr{n}= a\sbr{n}+g^{c}\sbr{n}+\alpha_{1,o}\frac{d^{2}\delta a\sbr{n}}{dt^{2}}+\alpha_{2,o}\delta g\sbr{n}.
\end{equation}

We have then prepared two data series to be used for the fit and the noise decomposition.

The first is  the double derivative  $\frac{d^2 \delta a}{d t^{2}}\sbr{n}$ of $\delta a\sbr{n}$. The second is   a copy $\delta g\sbr{n,\tau}$ of $\delta g\sbr{n}$ but delayed by an amount $\tau$.

A fit to  $g\sbr{n}$ was then performed in the frequency domain  with the fitting function  
\begin{equation}
\label{ff}
g^{fit}\sbr{n}=\alpha_{1}\frac{d^2 \delta a}{d t^{2}}\sbr{n}+\alpha_{2}\delta g\sbr{n,\tau}
\end{equation}
with $\alpha_1, \alpha_2,$ and $\tau$  being the  free parameters of the fit. 

All data series have been prepared with a sampling time of 10Hz, and then, after having applied the appropriate delays, low-passed and decimated to 1 Hz. In analogy to the LPF data analysis,  we further low-passed and decimated the data to 0.1 Hz.  We only fit the DFT coefficients with frequencies $f \le$ 50 mHz.

Furthermore  we have used  $N_{s}=9$ data stretches with 50\% overlap, and a Blackman-Harris spectral window with $k_{o}=4$. Finally we have used $k_{1}=1$.

\begin{figure}[h]
\begin{center}
\includegraphics[width=0.48\textwidth]{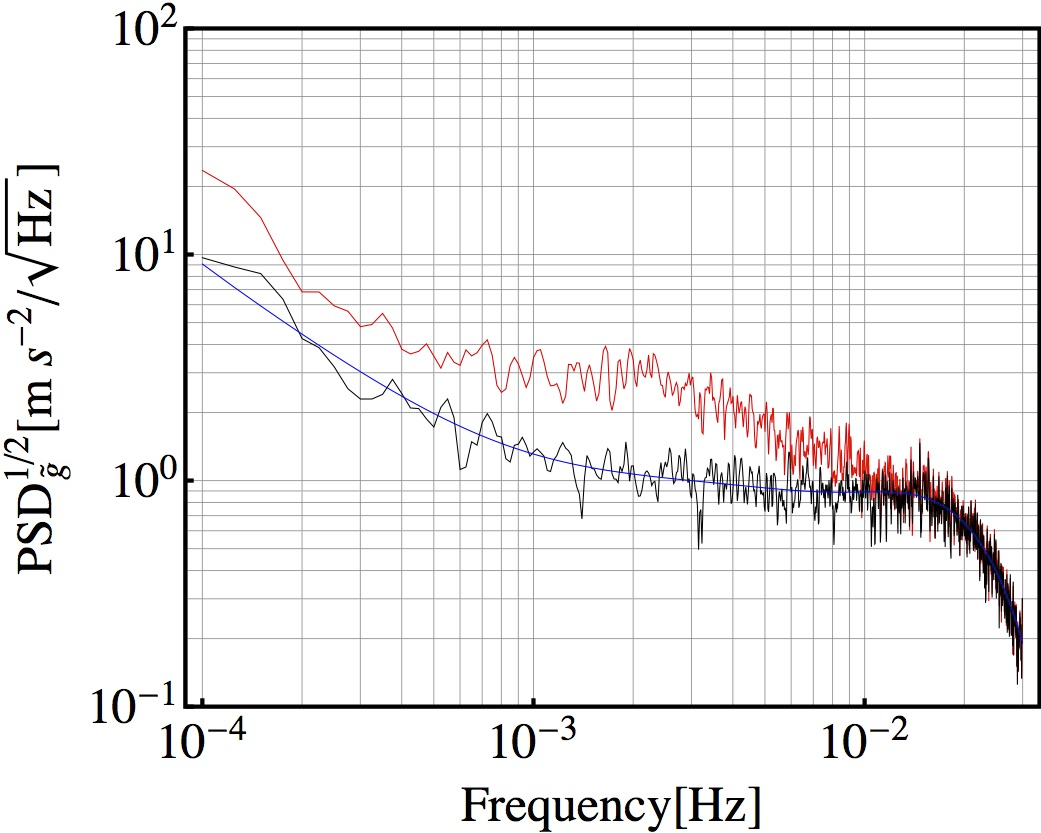}
\end{center}
\caption{Result of noise projection with simulated data. Upper noisy red line: the square root  PSD of the base data series $g\sbr{n}$. Lower noisy black line: the square root PSD of the residuals after fitting $g\sbr{n}$, with  $\alpha_{1}\frac{d^2 \delta a}{d t^{2}}\sbr{n}+\alpha_{2}\delta g\sbr{n,\tau}$. The solid blue line represents the theoretical value of the expected PSD.}

\label{master plot}   

\end{figure}

Fig. \ref{master plot} shows a typical  result of the fit.

\begin{figure*}
\begin{center}
\includegraphics[width=0.9\textwidth]{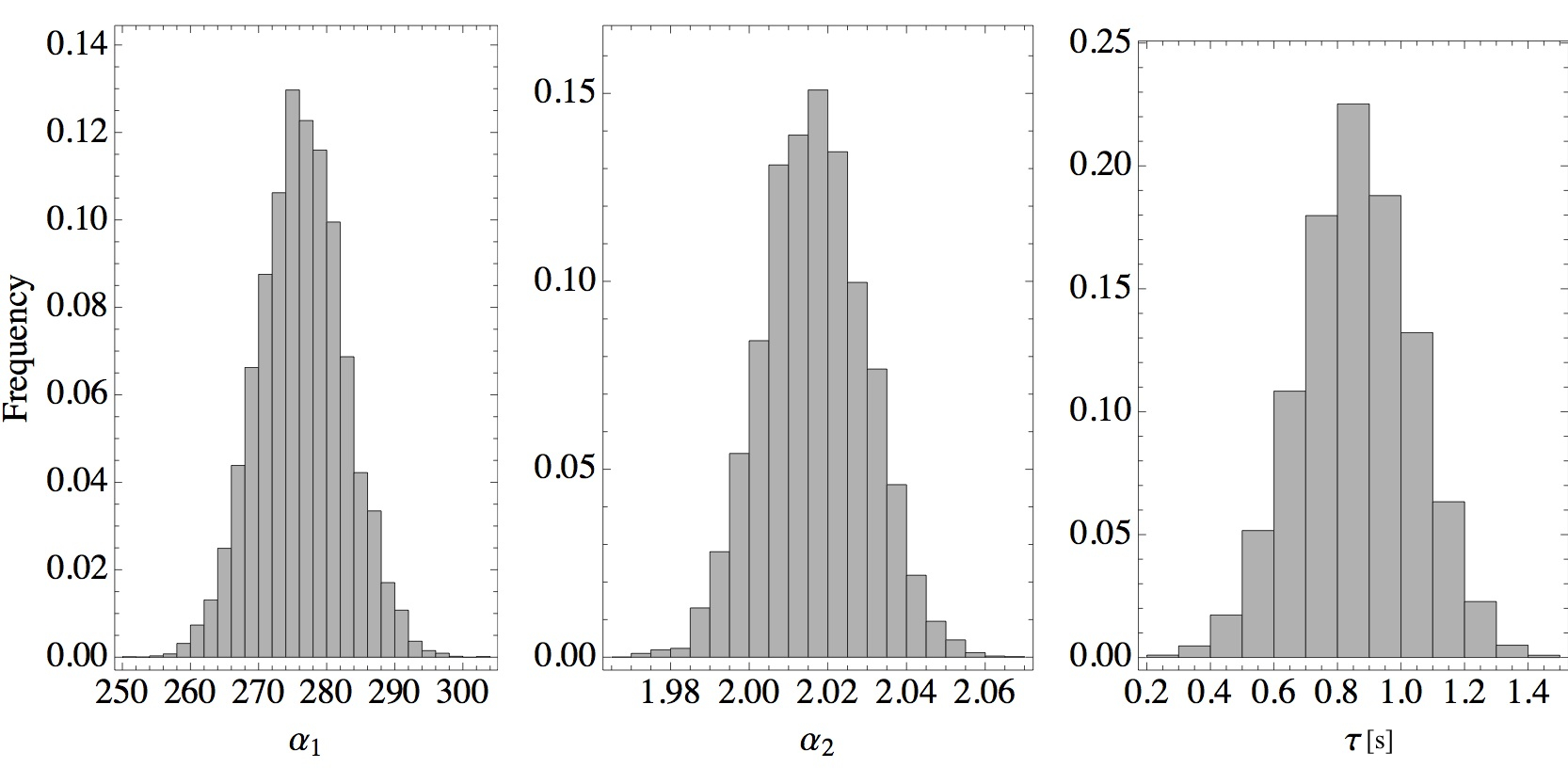}
\end{center}
\caption{Marginalized histograms for the three fitting parameters, obtained with the MCMC method and the LL.}

\label{histograms}   

\end{figure*}

The fit has been performed with  the MCMC method  and the Metropolis-Hastings algorithm with the LL. The parameter values used for the plot  in Fig. \ref{master plot} correspond to the maximum of the likelihood for that specific realisation of the data. 
An example of the marginalized MCMC distributions of the different fitting parameters is reported in Fig. \ref{histograms}.

The simulation has been repeated $N_{rep}=40$ times. Each time we have generated a new data series  to which we have applied the  MCMC fit. For each repetition we have recorded the mean values  and the standard deviations of the MCMC distributions of the three fitting parameters. 

For comparison we have also performed  similar  simulations  for the following cases:

\begin{itemize}
\item $k_{1}=4$ instead of $k_{1}=1$. 
\item $k_{1}=1$ and $\gamma=1/2$.
\item $k_{1}=1$ and $\gamma=1/3$. 
\item { $k_{1}=1$, but with the IRLS method. In this case, in each simulation, the IRLS fit has been performed with a numerical minimisation routine  until a stationary solution has been reached. Then an MCMC sequence has been generated by using the likelihood in Eq. \ref{eq:lsalt}, with the $S_{k}$ equal to residuals of the stationary solution.}
\end{itemize}

The results are summarised in Table \ref{tab:N1}.
 
\begin{table*}

\begin{center}
\caption{\\
Results from  simulations. The meaning of the columns is the following. True: parameter values used in the simulation. Columns under the header  `LL'  refer to fits performed with the LL method. Columns under the header `IRLS' refer to fits performed with the IRLS method. Av.: average of  the 40 mean values of the  MCMC distributions obtained in the 40 independent simulations. $\sigma_\text{sample}$: standard deviation of the 40 mean values. $\sigma_\text{MCMC}$:  root mean square  of the  standard deviations of the MCMC distributions  of the 40 independent simulations.}
\label{tab:N1}
\vspace*{1\baselineskip}
\scalebox{.85}{\noindent\(\begin{array}{| c  c | c  c  c | c  c  c  | c  c c | c  c  c  | c  c c |}
\multicolumn{2}{c}{}& \multicolumn{12}{c}{LL}& \multicolumn{3}{c}{\text{IRLS}}\\
\cline{4-13} \cline{15-17} 
\multicolumn{2}{c}{}& \multicolumn{3}{c}{  k_{1}=1, \gamma=1}& \multicolumn{3}{c}{ k_{1}=4,  \gamma=1}& \multicolumn{3}{c}{ k_{1}=1,  \gamma=1/2}& \multicolumn{3}{c}{ k_{1}=1,  \gamma=1/3}& \multicolumn{3}{c}{  k_{1}=1,  \gamma=1}\\
\hline 
 \text{Param.} & \text{True} & \text{Av.} & \sigma _{\text{sample}} & \sigma _{\text{MCMC}} & \text{Av.} & \sigma _{\text{sample}} & \sigma _{\text{MCMC}} & \text{Av.} & \sigma _{\text{sample}} & \sigma _{\text{MCMC}}& \text{Av.} & \sigma _{\text{sample}} & \sigma _{\text{MCMC}}& \text{Av.} & \sigma _{\text{sample}} & \sigma _{\text{MCMC}} \\
\hline
 \alpha_1 & 300 & 300 & 7.9 & 5.8& 300 & 12 & 12 &302&10&8.4&301&8.9&10& 297& 9.8 & 5.5\\
\alpha_2 & 2 & 2.005 & 0.026 & 0.014&2.004 & 0.029 & 0.027&2.005&0.021&0.019&2.001&0.026&0.023&2.000 & 0.025 & 0.013 \\
 \tau\left[s\right]  & 0.8 & 0.84 & 0.29 & 0.18&0.70 & 0.42 & 0.36&0.79&0.34&0.26&0.77&0.37&0.31&0.90 & 0.33 & 0.17 \\
 \hline
\end{array}\)}
\end{center}
\end{table*}

It is worth pointing out the following facts that can be observed in Table \ref{tab:N1}:
\begin{itemize}
\item For all fitting parameters, the average values resulting from the different fitting methods, agree with each other and, with the true values, within their respective errors $\sigma_{sample}/\sqrt{N_{rep}}$, with the exception of the results of IRLS simulation. These results  however agree with the rest at worst to within $1.8 \sigma_{sample}/\sqrt{N_{rep}}$ . In particular this confirms that the LL method produces an unbiased estimate of the parameters.
\item On the contrary, for  $\gamma=1$,  $\sigma_{sample}$ and $\sigma_{MCMC}$ agree for all parameters, within their relative uncertainties,  \emph{only for the case $k_{1}=4$}. In this last case however, both $\sigma_{sample}$ and $\sigma_{MCMC}$ are somewhat larger than those obtained with $k_{1}=1$, both for case of the LL and for that of  IRLS.
\item The agreement between $\sigma_{sample}$ and $\sigma_{MCMC}$ may be recovered also for $k_{1}=1$, if $\frac{1}{3} < \gamma < \frac{1}{2}$.
\item The IRLS procedure followed by the MCMC likelihood map, takes approximately 4 times the time needed by the  straightforward LL, MCMC map.
\end{itemize}

\subsection{\label{LPF} Application to LPF simulator data analysis}

As is customary with space missions, an end-to-end mission simulator of LPF has been set up by industry\cite{simulator}. The simulator includes not only the linear dynamics of satellite and TMs translation, but also realistic,  non-linear models of the critical parts of the system, like  the  electrostatic actuation system, the rotational dynamics of  both the TMs and the satellite, the interferometers, etc. These non linearities are not expected  to play a significant role, as all displacements, velocities, etc. are expected to be very small during science operations. We expect then to be able to understand the largest part of LPF results within a linear model, and to have to deal with non linearities only as occasional small deviations from the linear regime. The simulator is proving extremely useful for testing this approach and the related  data analysis algorithms and tools \cite{ltpda}. 

In this section we illustrate the application of the method described in Sec. \ref{sec:2}, to  some selected  cases, taken from  the extensive simulation campaigns that have been performed in preparation for the mission operations. Specifically, we first discuss  a simulated instrument calibration, where large calibration guidance signals are injected in the proper control loops. We then give an example of true noise decomposition performed to hunt for the source of some extra noise found in the  simulated data.  

\subsubsection{\label{calibration} Extraction of calibration signals}

As explained in Sec. \ref{newintro},  TM2 is forced, by a weak electrostatic force control loop driven by the main interferometer signal $s_{12}$, to stay nominally at a fixed distance from the reference TM.

This loop compensates most of the low frequency ($< 10\,\rm{mHz}$)  forces $g$ by applying commanded forces $g^c$. Thus, accurate subtraction of these applied forces, including calibration of the actuator, is needed to extract the disturbance forces. Our analysis subtracts the commanded force series multiplied by a calibration factor, $A g_{2}^c$.  $A$ is extracted with a dedicated ``system identification'' experiment in which a comparatively large modulated guidance signal is added to the measured displacement signal, $s_{12}$, in the control loop. The injected signal has the effect of modulating the distance between the TMs by exerting forces much larger that those exerted by the loop in the presence of just the background  noise.

Our technique to calibrate the transfer function consists of  fitting the  pre-processed commanded force series  $A g_{2}^{c}\sbr{n,\vec{\theta}}$ to $a\sbr{n}$. As already mentioned, by `pre-processed' we mean here that the series are  filtered via an algorithm that depends on some set of parameters, $\vec{\theta}$, that  become free fitting parameters. Commonly  for the $g_{2}^{c}\sbr{n,\vec{\theta}}$ series, $\vec{\theta}$ only includes the amplitude, $A$, and a delay, $\tau$, but single pole filters that simulate the response of the actuators have also been tested in the past \cite{ltpda, oldMCMC, Congedo2012, FerraioliMCMC}.
In addition to the feedback force removal, we also subtract, from the acceleration data, the forces due to the motion of both TMs within the static force gradient in the satellite (see Eq. \ref{dynamics}). This is  predicted to be dominated by the electric gradient due to  various voltages applied between the TMs and their surroundings, and by the gravitational gradient. As already explained, these forces are proportional to the TM displacements relative to the satellite, measured by $s_{1}$ and $s_{12}$.  In the calibration experiment these displacements are also large, so that the effect of the respective  forces may have appreciable SNR. As a consequence, we include  in the fitting model  two terms proportional to  $s_{1}$ and to $s_{12}$, respectively.  In conclusion we fit the acceleration data series assuming the following model

\begin{equation}
\label{eq:3.1}
\begin{split}
& \tilde{a}\left[k\right]=A  \tilde{g}_{2}^{c}\left[k,\tau \right]-\omega_{2}^{2} \tilde{s}_{12}\left[k\right]\\
& -\left(\omega_{1}^{2} - \omega_{2}^{2}\right) \tilde{s}_{1}\left[k\right]+\tilde{g}\left[ k \right],\\
\end{split}
\end{equation}
with the amplitudes $A$, $\omega_{2}^{2}$, $ \left(\omega_{1}^{2} - \omega_{2}^{2}\right)$, and the delay $\tau$ as free fitting parameters.

For all the fits, we have used the Blackman-Harris spectral window,  taken $N_{s}=5$ and limited the frequency range to $f \le$ 50 mHz.

\begin{figure}[h]
\begin{center}
\includegraphics[width=0.48\textwidth]{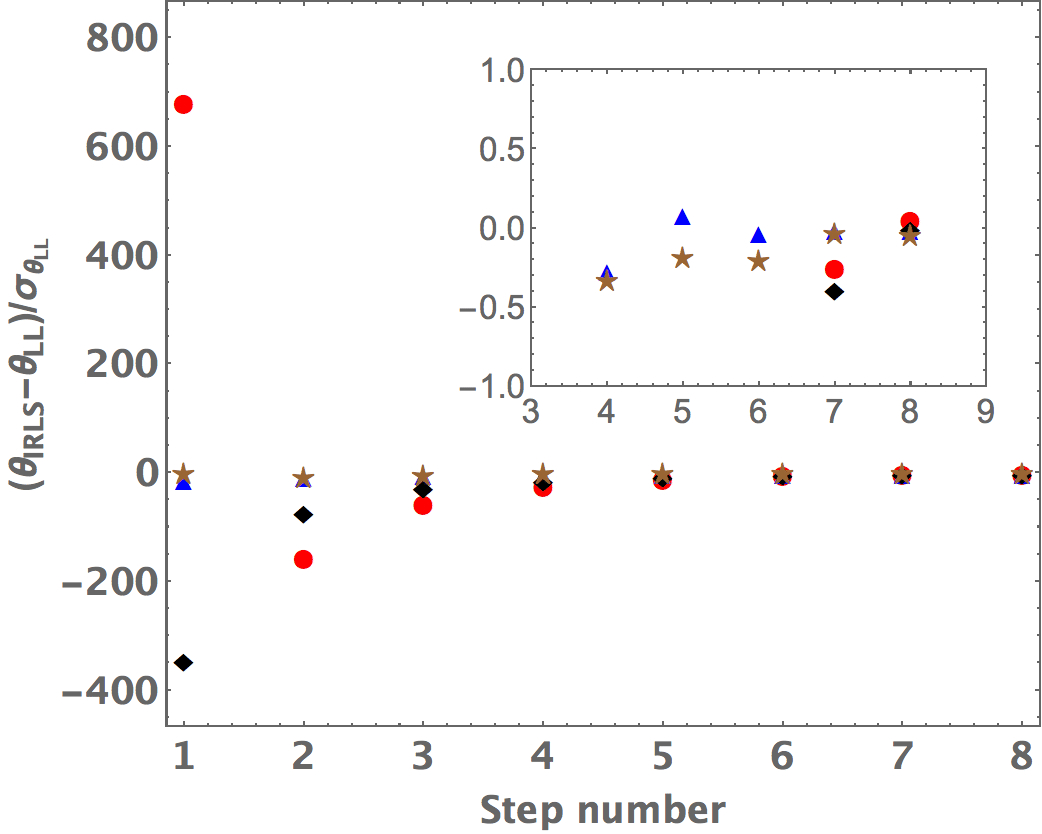}
\end{center}
\caption{  Deviation of the best fit parameter  obtained at each step of the IRLS procedure, from those obtained with the MCMC mapping of the LL. Deviations are divided by the standard deviation obtained from the MCMC chain of the LL. Colours mark the different parameters. Red circles: A.  Blue triangles: $-\omega_{2}^{2} $. Black diamonds: $ -\left(\omega_{1}^{2} - \omega_{2}^{2}\right)$. Brown stars: $\tau$. } 
\label{fig:convergence}   

\end{figure}

For the sake of comparison, the fit has been performed with  both  the  LL and the IRLS methods. In this last case, the convergence of the best-fit parameter values to those obtained with the LL method has been verified and the results are reported in Fig. \ref{fig:convergence}.

\begin{figure}[h]
\begin{center}
\includegraphics[width=0.48\textwidth]{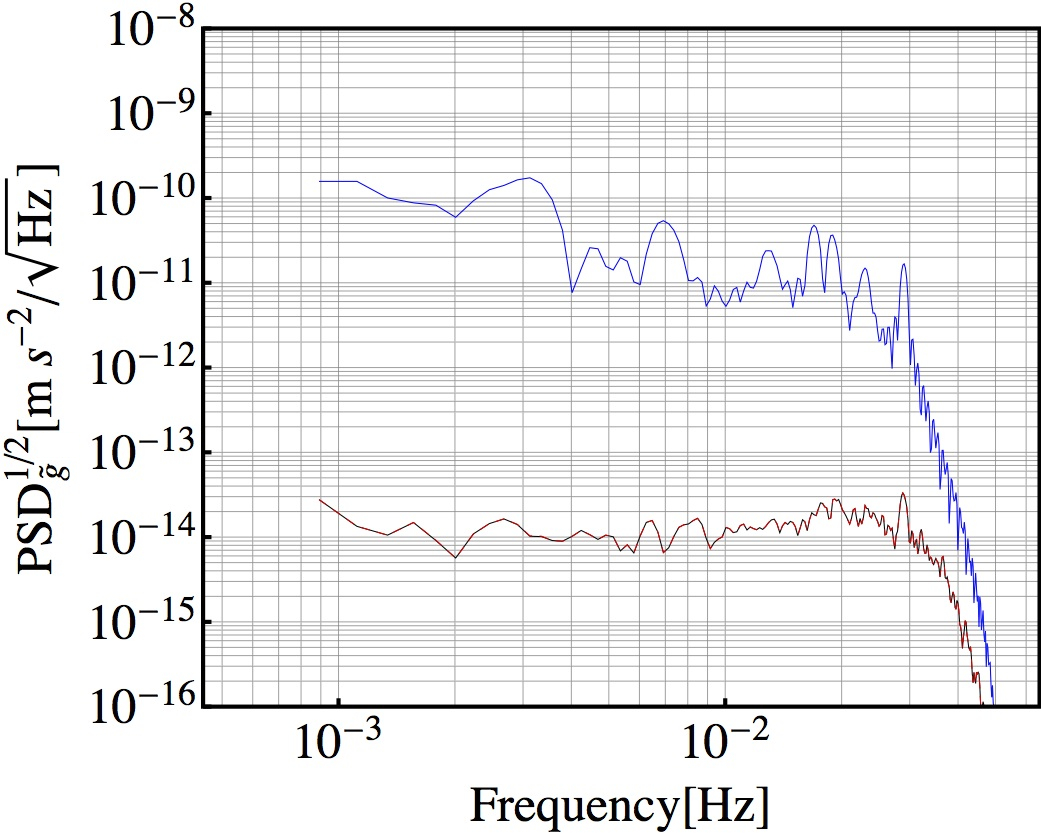}
\end{center}
\caption{ PSD of various data series. The  upper, blue, noisy continuous line is the PSD of  $ a\sbr{n}$ \emph{before the fit}. The lower, black, noisy continuous line, barely visible behind the red dashed line, is the PSD of the residuals of the fit with the LL. The red dashed line is the PSD of the residuals of the IRLS fit. } 
\label{fig:3045}   

\end{figure}

Furthermore, Fig. \ref{fig:3045} shows the PSD of the residuals of the two different  fits.

Finally, in Table \ref{tab:3.1} we report, for the same fits, the parameter values and their uncertainties.

\begin{table*}
\begin{center}
\caption{\\
Parameter values from the various fits of Fig. \ref{fig:3045}. `LL' refers to the fit performed with the LL. `IRLS' to the fit done with the IRLS method. `Mean' indicates  the mean of the MCMC parameter distribution. $\sigma$ represents its standard deviation.}

\label{tab:3.1}
 \vspace*{1\baselineskip}
\scalebox{.85}{\noindent\(\begin{array}{| c | c c | c c |}
\hline

\multicolumn{1}{| c }{Parameter}& \multicolumn{2}{| c |}{\text{LL}} & \multicolumn{2}{| c |}{\text{IRLS}}\\

 & \text{Mean} & \text{$\sigma$}  & \text{Mean} & \text{$\sigma$} \\
\hline
 -\omega _2^2\left[s^{-2}\right] & 2.259\times 10^{-6} & 1\times 10^{-9} & 2.259\times 10^{-6} & 1\times 10^{-9} \\
 \left(\omega _2^2-\omega _1^2\right) \left[s^{-2}\right]& 7.2\times 10^{-7} & 1.2\times 10^{-7} & 7.1\times 10^{-7} & 1.2\times 10^{-7} \\
 \text{A} & 1.04998 & 1.2\times 10^{-5} & 1.04998 & 1.2\times 10^{-5} \\
 \tau\left[s\right]  & -4.002\times 10^{-1} & 3\times 10^{-4} & -4.002\times 10^{-1} & 3\times 10^{-4}  \\
\hline
\end{array}\)}
\end{center}
\end{table*}

In the case of the IRLS fit, as before, we have first run the re-weighted iteration, performing, at each step, just a numerical maximisation of the likelihood in Eq. \ref{eq:lsalt}. After obtaining the stationary solution for the $S_{k}$, we have inserted these values  back in the likelihood of Eq. \ref{eq:lsalt} and  performed  an  MCMC  map to estimate the parameter errors reported in \ref{tab:3.1}.
As expected, the results of the two methods are indistinguishable.

It must be stressed that, unfortunately,  we cannot straightforwardly compare the results of Table \ref{tab:3.1} with the ``true'' parameter values used in the simulator. Only the parameter A=1.05 has a fixed and traceable value, while both $\omega _2^2$ and $\omega _1^2$ are heuristic single-parameter approximations of a complicated model where the electrical fields, one of the major sources of force gradient, are dynamically calculated as part of the overall three body dynamics of the TMs and the satellite. This is also the case for $\tau$ that results from the combination of various  propagation delays within the system. All that said, the values resulting from the fit agree with the expectation, within their respective uncertainties.

\subsubsection{\label{mystery noise} Decomposition of low frequency noise}

The second case of study deals with  a typical case of noise hunting and decomposition. We started by forming a time series for  $g\sbr{n}$, subtracting the commanded force and stiffness effects as calibrated in the previous section. The PSD of $g\sbr{n}$ at frequencies  below 1 mHz was found in excess of what was expected from the physical model at the basis of the simulator. However calculating the expected PSD  is not straightforward, given the complexity of the simulator model, so that no firm conclusion could be made about  this noise being  a real property of the  system or rather a software artefact.

\begin{figure}[h]
\begin{center}
\includegraphics[width=0.48\textwidth]{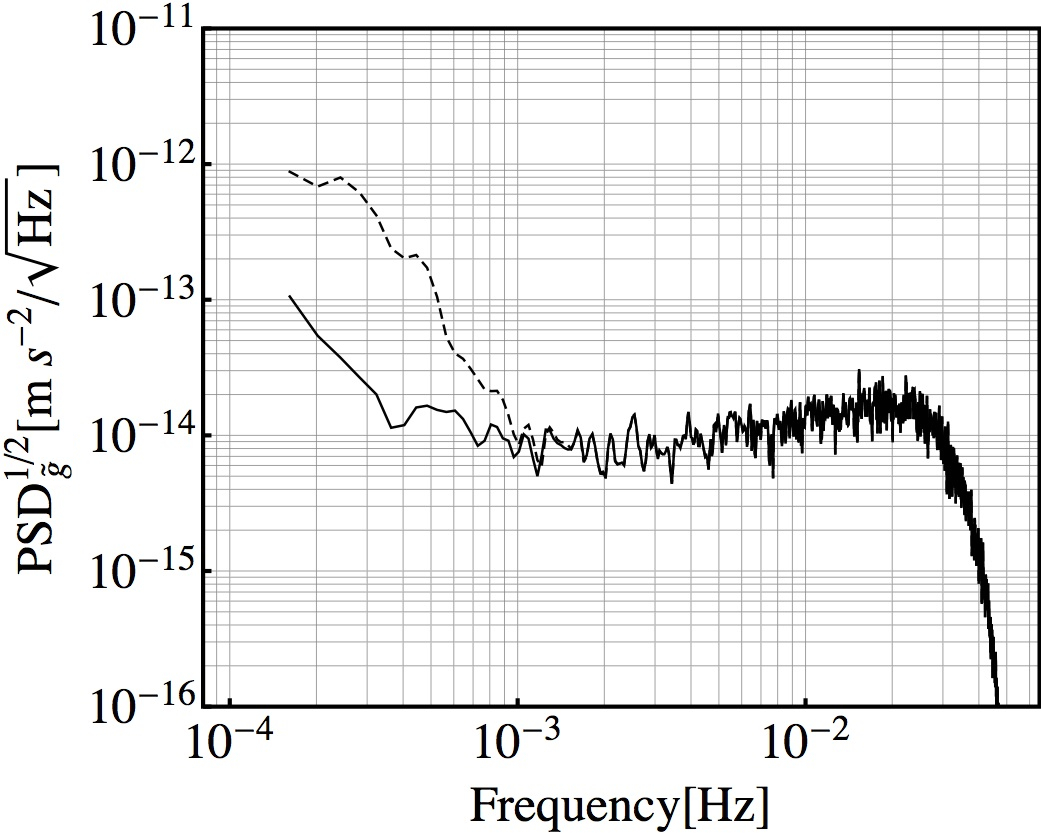}
\end{center}
\caption{The effect of subtraction of $\Delta F_{z}$ ,  $\Delta F_{y}$, and $N_{x}$. The dotted line represents the PSD of the data series before subtraction. The solid line represents the PSD of the data series after subtraction, i.e. the series of the fit residuals.}

\label{fig:mistery}   

\end{figure}

To help in identifying the source, we performed an extensive campaign of decomposition, fitting all available data series to $g\sbr{n}$  in an attempt to identify the source.  Not knowing what to expect as a background noise, we were forced to develop the methods we are discussing here. Only three data series were observed  to  significantly reduce the noise when subtracted from  $g\sbr{n}$. The first and most important series was the difference $\Delta F_{z}$ between the forces applied by the control system to the two different TMs along the $z$-axis (see Fig. \ref{lpf} for the definition of the axes). We also found a smaller but significant contribution of   $\Delta F_{y}$, the corresponding difference of force  along $y$. We finally found another small contribution from the torque $N_{x}$ applied to the inertial reference TM along the $x$-axis. Fig. \ref{fig:mistery} shows the effect of subtraction.

In the simulator model, these forces should leak into $x$ via some linear coupling coefficients, the values of which are known simulator input parameters. We found that the values  of the coupling coefficients resulting from the fit  are in quantitative agreement with those in the simulator, for  $\Delta F_{y}$ and $N_{x}$ (see Table \ref{tab:mystery}). 
\begin{table}
\begin{center}
\caption{Parameter values from  LL noise decomposition. Mean and $\sigma$ are the mean and standard deviation from the MCMC  chain. ``Expected'' refer to the values in the simulator.}
\label{tab:mystery}
\vspace*{1\baselineskip}
\scalebox{.85}{\noindent\(\begin{array}{| c | c | c | c |}
\hline
\multicolumn{1}{| c |}{\text{Disturbance}}& \multicolumn{3}{| c |}{\text{Coupling coefficients}}\\
\cline{1-1} \cline{2-3} 
\hline 
 \text{} & \text{Mean}& \sigma  &\text{Expected} \\
\hline
 \Delta F_{z} & -8.03 \times 10^{-3} & 6 \times 10^{-5} & 1.1 \times 10^{-3} \\
\Delta F_{y} & 1.3 \times 10^{-3} & 2 \times 10^{-4} & 1.1 \times  10^{-3}\\
 N_{x}& 7.9 \times 10^{-5}  m^{-1} & 9 \times 10^{-6} m^{-1} & 7.7 \times 10^{-5}  m^{-1}\\
\hline
\end{array}\)}
\end{center}
\end{table}
The Table also shows that,  on the contrary, the coupling coefficient for $\Delta F_{z}$ is $\sim 8$ times larger than the corresponding one used  in the simulator and, in addition also has the wrong sign.

The named forces and torques are commanded by a controller in charge of stabilising the absolute orientation of the satellite. Explaining how this controller works goes beyond the scope of the present paper. For the sake of the discussion it is only useful to mention that the  controller is driven by the signals from a set of Autonomous Star Trackers, so that our finding allowed us to trace the problem back to an erroneous coupling of these devices.

\section{\label{disc}Discussion and conclusions}

The results of the tests reported  in Secs. \ref{simul} and \ref{calibration} show that:
\begin{itemize}
\item{The PSD of the residuals of the fit  is in quantitative agreement with the expected spectrum for the background noise.}
\item{The LL is well behaved and produces unbiased  results when used in a MCMC fit.}
\item{The  LL  MCMC fit, and the IRLS fit followed by a MCMC likelihood mapping, give the same results, this last one being substantially slower in the case of non linear fitting.}
\item{The estimate of the parameter errors, from the fit performed with $k_{1}=1$, seems indeed  to be moderately affected by the correlation between nearby DFT coefficients. This bias, as expected, is common   both to LL and IRLS fitting.}
\item {The bias can indeed be made negligible by taking properly spaced coefficients, or by correcting the likelihood with the proper factor $\gamma$. }
\end{itemize}

We believe then that the LL fitting presented here is of general use and solves the problem of the lack of {\it a priori} knowledge of the target noise in frequency domain fits.
From the point of view of the computational load, for multi parameter nonlinear fitting, the method is faster than IRLS. However the method  is intrinsically nonlinear, and fitting would not reduce to a set of algebraic equations, as  IRLS does, when the  fitting parameters are just multiplicative  amplitudes. Given that the final results of the two method coincide, the IRLS method is preferable in the linear model case.

We think that the LL approach is numerically lighter than the  methods\,\cite{littenberg, Cornish0} that employ a parametric model for the noise in the likelihood in eq. \ref{eq:Gaussian probability}, and then sample the likelihood with MCMC over a parameter space that also include the noise model parameters. Indeed,  with that method the likelihood function contains more terms, the sum of the squares and the sum of the logarithm of the $S_{k}$, and the parameter space to be searched numerically is wider.  

We think that the method used here, including the partitioning of the data   in stretches and the average over the stretches, could be usefully extended also to case of signal extraction from GW detectors.  This is particularly true in the case of a space borne detector like eLISA, which is expected to be signal dominated at all times, so that a direct measurement of the instrument noise is difficult. An unbiased estimator of the noise could then help in avoiding the introduction of unwanted bias in the signal parameter estimation.
\section{\label{Appendix} Appendix: properties of DFT coefficients}
It is a well-known result \cite{percival} that
\begin{equation}
\label{eq:A1.2}
\begin{split}
&\langle \abs{ \tilde{g}\left[k\right] }^{2} \rangle = S_{k}.
\end{split}
\end{equation}
With the same kind of calculations that lead to eq. \ref{eq:A1.2}, one can easily calculate that:
\begin{equation}
\label{eq:A1.3}
\begin{split}
&\langle {Re\lbrace \tilde{g}\left[k\right] \rbrace}^{2} \rangle =\frac{1}{2\pi}\int_{-\pi }^{\pi }d\phi~\tilde{S}_{\tilde{g}}\left(\phi\right)\frac{1}{4} \times\\ &\times \Biggl( ~ {\abs{w\left(\phi-k\frac{2\pi}{N}\right)}^{2} +\abs{w\left(\phi+k\frac{2\pi}{N}\right)}^{2} }+\\&+2 Re \set{w\left(\phi+k\frac{2\pi}{N}\right)w^{*}\left(\phi-k\frac{2\pi}{N}\right)}  \Biggr),\\
\end{split}
\end{equation}
\begin{equation}
\label{eq:A1.32}
\begin{split}
&\langle {Im\lbrace \tilde{g}\left[k\right] \rbrace}^{2} \rangle =\frac{1}{2\pi}\int_{-\pi }^{\pi }\tilde{S}_{\tilde{g}}\left(\phi\right)  \frac{1}{4}\times\\&\times\Biggl(~ {\abs{w\left(\phi-k\frac{2\pi}{N}\right)}^{2} +\abs{w\left(\phi+k\frac{2\pi}{N}\right)}^{2} }+\\&- 2 Re \set{w\left(\phi+k\frac{2\pi}{N}\right)w^{*}\left(\phi-k\frac{2\pi}{N}\right)}  \Biggr)d\phi.\\
\end{split}
\end{equation}
The consequence of Eqs. \ref{eq:A1.3} and \ref{eq:A1.32}  is that if $k$ is large enough so that $w\left(\phi-k\frac{2\pi}{N}\right)$ has no overlap with $w\left(\phi+k\frac{2\pi}{N}\right)$, then 
\begin{equation}
\label{eq:A1.5}
\begin{split}
&\langle {Re\lbrace \tilde{g}\left[k\right] \rbrace}^{2} \rangle =\langle {Im\lbrace \tilde{g}\left[k\right] \rbrace}^{2} \rangle =\frac{1}{2}\langle \abs{ \tilde{g}\left[k\right] }^{2} \rangle = \frac{1}{2}S_{k}.
\end{split}
\end{equation}
In addition one can calculate that:
\begin{equation}
\label{eq:A1.4}
\begin{split}
&\langle {Re\lbrace \tilde{g}\left[k\right] \rbrace} {Im\lbrace \tilde{g}\left[k\right] \rbrace} \rangle =\\&\frac{1}{2\pi}\int_{-\pi }^{\pi }d\phi \tilde{S}_{\tilde{g}}\left(\phi\right) \frac{1}{2} Im \set{ w\left(\phi+k\frac{2\pi}{N}\right) w^{*}\left(\phi-k\frac{2\pi}{N}\right)}.
\end{split}
\end{equation}
So that again, if $k$ is large enough, then

\begin{equation}
\label{eq:A1.4b}
\begin{split}
&\langle {Re\lbrace \tilde{g}\left[k\right] \rbrace} {Im\lbrace \tilde{g}\left[k\right] \rbrace} \rangle =0.
\end{split}
\end{equation}
Finally it is also straightforward to get that 

\begin{equation}
\label{eq:A1.6}
\begin{split}
&\langle { \tilde{g}\left[k\right]} { {\tilde{g}}^{*}\left[k'\right]} \rangle =\frac{1}{2\pi}\int_{-\pi }^{\pi }d\phi\tilde{S}_{\tilde{g}}\left(\phi\right)\times\\&\times w^{*}\left(\phi-k\frac{2\pi}{N}\right) w\left(\phi-k'\frac{2\pi}{N}\right),
\end{split}
\end{equation}

\begin{equation}
\label{eq:A1.62}
\begin{split}
&\langle { \tilde{g}\left[k\right]} { {\tilde{g}}\left[k'\right]} \rangle =\frac{1}{2\pi}\int_{-\pi }^{\pi }d\phi\tilde{S}_{\tilde{g}}\left(\phi\right)\times\\& \times w\left(\phi-k\frac{2\pi}{N}\right) w\left(\phi-k'\frac{2\pi}{N}\right). 
\end{split}
\end{equation}

Thus, if  $k$ and $k'$ are spaced enough that $w\left(\phi-k\frac{2\pi}{N}\right)$ has no overlap with $w\left(\phi-k'\frac{2\pi}{N}\right)$, then $\tilde{g}\left[k\right] $ and $\tilde{g}\left[k'\right] $ are independent random variables.

\begin{figure}
\begin{center}
 \includegraphics[scale=0.5]{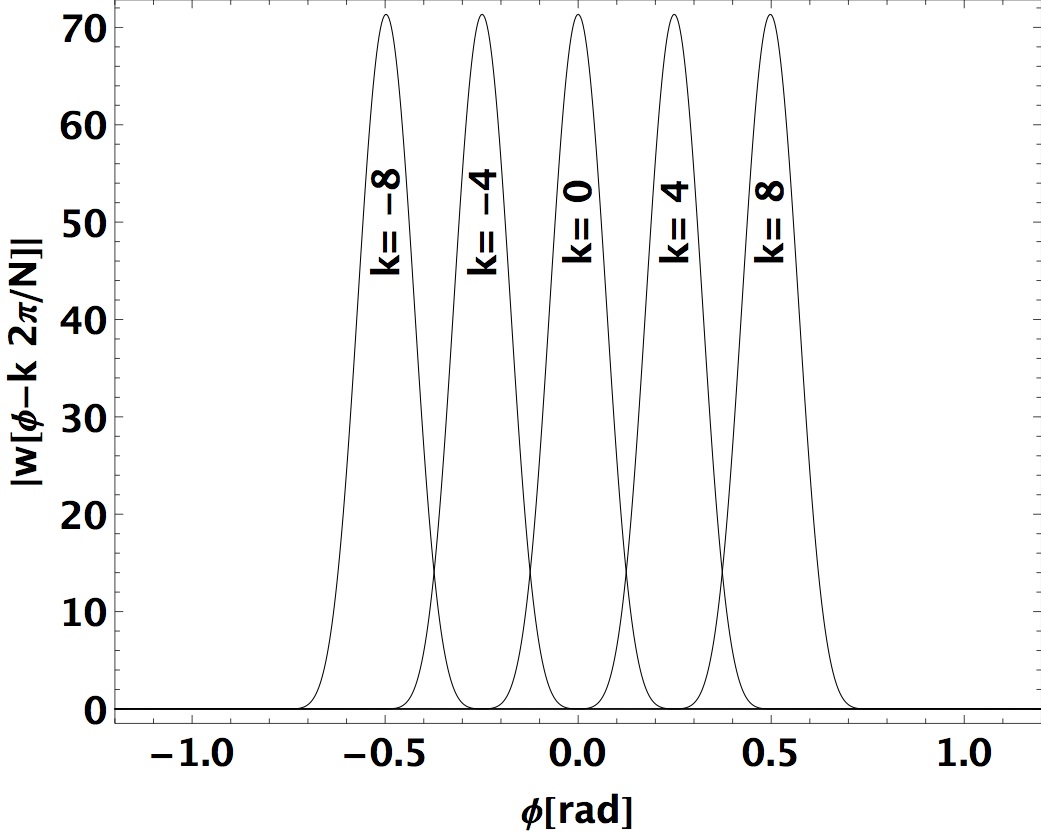}
\end{center}
\caption{The spectral window  for the Blackman-Harris plotted for different values of k, spaced by $k_{o}=8$}.

\label{fig:A1}   

\end{figure}

The amount of spacing needed for all the conditions above to hold depends on the selected spectral window $w\left[n\right]$. In general there exists a common value $k_{o}$ for which, with good accuracy, Eq. \ref{eq:A1.5} holds if $\abs{k}\ge \frac{k_{o}}{2}$, and $\tilde{g}\left[k\right] $ are independent random variables if $\abs{k-k'} \ge k_{o}$. With the Blackman-Harris window used in the numerical calculations of the present paper,  $k_{o}=8$ (see Fig. \ref{fig:A1}).

\begin{figure}
\begin{center}
\includegraphics[scale=0.45]{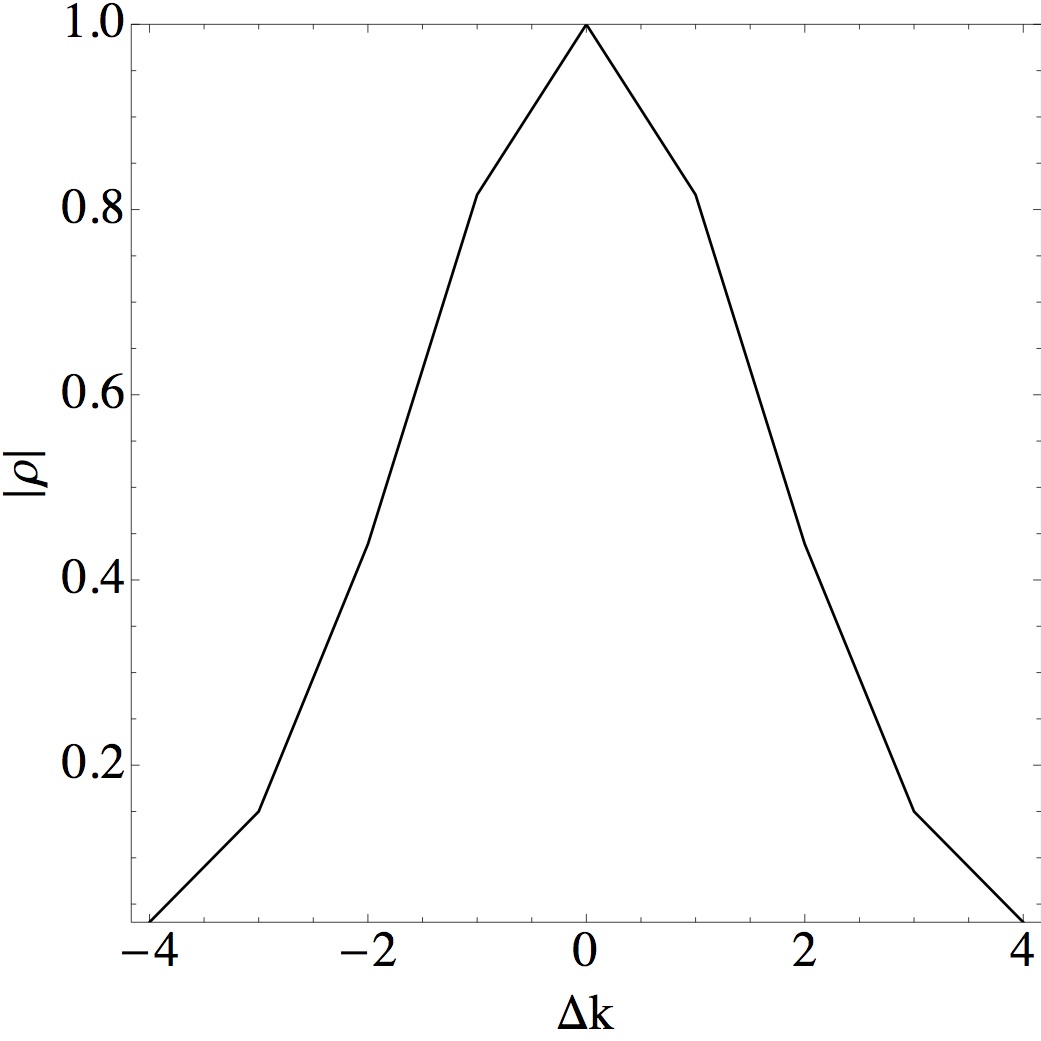}
\end{center}
\caption{The absolute value of the correlation coefficient defined  in eq. \ref{eq:A1.8},  as a function of the DFT coefficient spacing $\Delta k$. Values are calculated for   the Blackman-Harris window in the case of white noise.}
\label{fig:A2}   
\end{figure}

The correlation coefficients resulting from equation \ref{eq:A1.6}, may be calculated explicitly, for nearby DFT coefficients, assuming the noise is white. In figure \ref{fig:A2} we report the values for

\begin{equation}
\label{eq:A1.8}
\begin{split}
|\rho\left(\Delta k\right)|=&\frac{|\langle Re\lbrace \tilde{g}\left[k\right] \rbrace Re\lbrace { {\tilde{g}}\left[k+\Delta k\right]} \rbrace \rangle |}{\sqrt{\langle {|Re\lbrace \tilde{g}\left[k\right] \rbrace|}^{2}\rangle \langle {|Re\lbrace \tilde{g}\left[k+\Delta k\right] \rbrace|}^{2}\rangle}}=\\&=\frac{|\langle Im\lbrace \tilde{g}\left[k\right] \rbrace Im\lbrace { {\tilde{g}}\left[k+\Delta k\right]} \rbrace \rangle |}{\sqrt{\langle {|Im\lbrace \tilde{g}\left[k\right] \rbrace|}^{2}\rangle \langle {|Im\lbrace \tilde{g}\left[k+\Delta k\right] \rbrace|}^{2}\rangle}}.
 \end{split}
\end{equation}

For the first few  values of $\Delta k$. In addition one can calculate that 
\begin{equation}
\label{eq:A1.8b}
\begin{split}
&\frac{|\langle Re\lbrace \tilde{g}\left[k\right] \rbrace Im\lbrace { {\tilde{g}}\left[k+\Delta k\right]} \rbrace\rangle |}{\sqrt{\langle {|Im\lbrace \tilde{g}\left[k\right] \rbrace|}^{2}\rangle \langle {|Re\lbrace \tilde{g}\left[k+\Delta k\right] \rbrace|}^{2}\rangle}}\sim 0
 \end{split}
\end{equation}
Thus if noise does not vary too much over a range of order of $k_{o}$,  a reduced spacing among coefficients may also be considered. In the case of the Blackman-Harris window, Fig. \ref{fig:A2} suggests that coefficients could be taken every $\Delta k \sim 4 $ and still be treated as basically independent, and that even for $\Delta k \sim 2 -3 $ the effect of the correlation may still be negligible.
\newpage
\section{\label{Appendix2} Appendix: additional formulas}

To get the result is Eqs. \ref{eq:sands1}, and \ref{eq:sands2}, it is sufficient to substitute   $P\left(\vec{g}\middle \vert \vec{\theta},\vec{S}\right)$ taken from Eq. \ref{eq:averaged probability} into  Eq. \ref{eq:momenta} and perform the integral over $S_j$. We get:

\begin{widetext}
\begin{equation}
\label{eq:sands1A}
\begin{split}
&\left \langle S_{j} \middle \vert \overline{\vec{g}}  \right \rangle=\frac{N_{s}}{N_{s}-m-1}\frac{
\displaystyle\int_{}^{}\overline{{\left\vert\tilde{g}\sbr{j,\vec{\theta}}\right\vert}^{2}}\prod_{k \in Q}{\left({\overline{\left\vert\tilde{g}\sbr{k,\vec{\theta}}\right\vert^{-2}}}\right)}^{-\left(N_{s}-m\right)}P\left(\vec{\theta}\right)d\vec{\theta}}{\displaystyle\int_{}^{}\prod_{k \in Q}{\left({\overline{\left\vert\tilde{g}\sbr{k,\vec{\theta}}\right\vert^{-2}}}\right)}^{-\left(N_{s}-m\right)}P\left(\vec{\theta}\right)d\vec{\theta}}=\frac{N_{s}}{N_{s}-m-1}\left \langle \overline{{\left\vert\tilde{g}\sbr{j}\right\vert}^{2}}\right \rangle,\\
\end{split}
\end{equation}
\begin{equation}
\label{eq:sands2A}
\begin{split}
&\left \langle S_{j}^{2} \middle \vert \overline{\vec{g}}  \right \rangle=\frac{N_{s}^{2}}{\left(N_{s}-m-1\right)\left(N_{s}-m-2\right)}\frac{\displaystyle\int_{}^{}\overline{{\left\vert\tilde{g}\sbr{j,\vec{\theta}}\right\vert}^{4}}\prod_{k \in Q}{\left({\overline{\left\vert\tilde{g}\sbr{k,\vec{\theta}}\right\vert^{-2}}}\right)}^{-\left(N_{s}-m\right)}P\left(\vec{\theta}\right)d\vec{\theta}}{\displaystyle\int_{}^{}\prod_{k \in Q}{\left({\overline{\left\vert\tilde{g}\sbr{k,\vec{\theta}}\right\vert^{-2}}}\right)}^{-\left(N_{s}-m\right)}P\left(\vec{\theta}\right)d\vec{\theta}}=\\&=\frac{N_{s}^{2}}{\left(N_{s}-m-1\right)\left(N_{s}-m-2\right)}\left \langle \overline{{\left\vert\tilde{g}\sbr{j}\right\vert}^{4}}\right \rangle.\\
\end{split}
\end{equation}

\end{widetext}

\begin{acknowledgments}
This work has been supported in part under contracts from Agenzia Spaziale Italiana, Istituto Nazionale di Fisica Nucleare,  Deutsches Zentrum fŸr Luft- und Raumfahrt  e.V., Swiss Space Office, State Secretariat for Education, Research and Innovation, United Kingdom Space Agency, and Plan Nacional del Espacio of the Spanish Ministry of Science and Innovation.

\end{acknowledgments}

\bibliography{bibtest_2_0}

\end{document}